\documentclass[fleqn,usenatbib]{mnras}
\pdfoutput=1

\usepackage{newtxtext,newtxmath}

\usepackage[T1]{fontenc}

\DeclareRobustCommand{\VAN}[3]{#2}
\let\VANthebibliography\thebibliography
\def\thebibliography{\DeclareRobustCommand{\VAN}[3]{##3}\VANthebibliography}

\usepackage{graphicx}	
\usepackage{amsmath}	

\title[Chemo-dynamics and ages from seven Red Giants]{Chemo-dynamics and asteroseismic ages of seven metal-poor red giants from the Kepler field}

\author[Arthur Alencastro Puls et al.]{
Arthur Alencastro Puls,$^{1,5}$\thanks{E-mail: arthur.alencastropuls@anu.edu.au (AAP)}
Luca Casagrande,$^{1,5}$
Stephanie Monty,$^{1,5}$
David Yong,$^{1,5}$
Fan Liu,$^{2}$\newauthor
Dennis Stello,$^{3,4,5}$
Victor Aguirre B\o rsen-Koch,$^{4}$ and
Ken C. Freeman$^{1,5}$
\\
$^{1}$Research School of Astronomy and Astrophysics, Australian National University, Canberra, ACT 2611, Australia\\
$^{2}$Centre for Astrophysics and Supercomputing, Swinburne University of Technology, Melbourne, VIC 3122, Australia\\
$^{3}$School of Physics, University of New South Wales, Sydney, NSW 2052, Australia\\
$^{4}$Stellar Astrophysics Centre (SAC), Department of Physics and Astronomy, Aarhus University, Ny Munkegade 120, DK-8000 Aarhus C, Denmark\\
$^{5}$ARC Centre of Excellence for All Sky Astrophysics in 3 Dimensions (ASTRO3D), Australia\\
}

\date{Accepted 2021 November 30. Received 2021 November 14; in original form 2021 September 5}

\pubyear{2021}

\begin{document}
\label{firstpage}
\pagerange{\pageref{firstpage}--\pageref{lastpage}}
\maketitle

\begin{abstract}
In this work we combine information from solar-like oscillations, high-resolution spectroscopy and Gaia astrometry to derive stellar ages, chemical abundances and kinematics for a group of seven metal-poor Red Giants and characterise them in a multidimensional \emph{chrono-chemo-dynamical} space. Chemical abundance ratios were derived through classical spectroscopic analysis employing 1D LTE atmospheres on Keck/HIRES spectra. Stellar ages, masses and radii were calculated with grid-based modelling, taking advantage of availability of asteroseismic information from Kepler. The dynamical properties were determined with \texttt{Galpy} using Gaia EDR3 astrometric solutions. Our results suggest that underestimated parallax errors make the effect of Gaia parallaxes more important than different choices of model grid or -- in the case of stars ascending the RGB -- mass-loss prescription. Two of the stars in this study are identified as potentially evolved halo blue stragglers. Four objects are likely members of the accreted Milky Way halo, and their possible relationship with known accretion events is discussed.
\end{abstract}




\section{Introduction}
\label{sec:intro}

Since the first proposal of \citet{Searle:1978aa} the hierarchical formation of the Milky Way (hereafter, the Galaxy), supported by the $\Lambda$CDM model \citep{White:1978,Kobayashi:2011aa,Planck:2016}, has been an object of debate and a target of surveys \citep{Gilmore:2012,De-Silva:2015aa,Majewski:2017}. One avenue to trace the assembly history of our Galaxy is by using stars, and in particular their kinematics, chemical abundances, ages and spatial distributions.

Signatures of the formation and evolution of our Galaxy lie in the chemical compositions of stars \citep[e.g.,][]{Venn:2004aa}. One key abundance ratio is [$\alpha$/Fe], which provides a measure of the star formation history for the following reason. The $\alpha$ elements are primarily produced in massive stars that explode as Type II supernovae (SNe) on short timescales (10$^7$ years) whereas Fe is produced in Type Ia SNe on longer timescales ($\sim$~10$^9$ years) \citep[e.g.,][]{Tinsley:1979aa,Matteucci:1986aa,McWilliam:1997aa}. This abundance ratio has thus provided key constraints on the relative star formation histories of the thin disk, thick disk, halo, bulge and dwarf galaxies \citep[e.g.,][]{Freeman:2002aa,Tolstoy:2009aa}.

The history of our Galaxy can also be traced using stellar kinematics. 
After the second data release of the Gaia mission \citep{GaiaDR2:2018} it has become increasingly clear that a massive dwarf galaxy merged with our Galaxy at z $\sim$ 3, which is consistent with the hierarchical formation hypothesis from $\Lambda$CDM. The signature of that merger was only detected through the combination of precise distances and kinematics from Gaia and chemical abundances from spectroscopy \citep{Belokurov:2018,Haywood:2018,Myeong:2018,Helmi:2020}.

It is also clear that, in addition to stellar chemical abundances and kinematics, a more detailed understanding of the evolution of our Galaxy requires accurate stellar ages. Until recently, however, accurate stellar ages have been difficult to obtain \citep{Soderblom:2010,Chaplin:2011}. In the case of red giant stars, masses are a good proxy for ages, as most of the stellar lifetime is spent in the main sequence, before they evolve to the (relatively) short red giant phase \citep{Kippenhahn:2012}. Recently, exquisite photometric data collected by space-based missions such as Kepler \citep{Koch:2010}, K2 \citep{Howell:2014}, CoRoT \citep{Baglin:2006aa} and TESS \citep{Ricker:2014} have allowed us to derive stellar radii, masses and ages with precision of a few percent \citep{Casagrande:2016aa,Anders:2017,SilvaAguirre:2018,SilvaAguirre:2020,Zinn:2021b}. Red giant stars display solar-like oscillations in their power spectra, which are characterised by a Gaussian envelope centered at the frequency of maximum power $\nu_{\mathrm{max}}$ and peaks separated by the average frequency spacing $\Delta \nu$ between consecutive orders of same angular degree \citep{Chaplin:2013aa,Hekker:2017}. It can be shown that these asteroseismic quantities are proportional to fundamental stellar properties such as average density, effective temperature and surface gravity, and, assuming that these proportionality relations can be scaled to the solar values, they can be rewritten as \citep{Ulrich:1986aa,Brown:1991ab,Kjeldsen:1995},

\begin{equation}\label{eq:mass_sc}
    \frac{M}{M_{\odot}} \simeq \left ( \frac{\nu_{\mathrm{max}}}{\nu_{\mathrm{max,\odot}}} \right )^{3} \left ( \frac{\Delta\nu}{\Delta\nu_{\odot}} \right )^{-4} \left ( \frac{T_{\mathrm{eff}}}{T_{\mathrm{eff,\odot}}} \right )^{1.5},
\end{equation}

\noindent and,

\begin{equation}\label{eq:radius_sc}
    \frac{R}{R_{\odot}} \simeq \left ( \frac{\nu_{\mathrm{max}}}{\nu_{\mathrm{max,\odot}}} \right )
\left ( \frac{\Delta\nu}{\Delta\nu_{\odot}} \right )^{-2} \left ( \frac{T_{\mathrm{eff}}}{T_{\mathrm{eff,\odot}}} \right )^{0.5},
\end{equation}

\noindent and also,

\begin{equation}\label{eq:logg_sc}
    \mathrm{log}\: g \simeq \mathrm{log}\: \left ( \frac{\nu_{\mathrm{max}}}{\nu_{\mathrm{max,\odot}}} \right ) + 0.5\:\mathrm{log} \left ( \frac{T_{\mathrm{eff}}}{T_{\mathrm{eff,\odot}}} \right ) + \mathrm{log}\: g_{\odot},
\end{equation}

\noindent where $M$ and $R$ are, respectively, stellar mass and radius, $T_{\mathrm{eff}}$ is the effective temperature, and $g$ is the surface gravity. Recent studies have shown that, in the case of stars ascending the Red Giant Branch (RGB), Eqs.~(\ref{eq:mass_sc}) and (\ref{eq:radius_sc}) tend to overestimate masses and, by a smaller extent, also radii, and a small non-linear correction would be required for them \citep[e.g.,][]{Miglio:2016aa,Gaulme:2016,Rodrigues:2017,Brogaard:2018}, however the results from \citet[][]{Zinn:2019apj885,Zinn:2020} are in agreement with radii derived from Gaia DR2 observations \citep[][]{GaiaDR2:2018} at the 2\% level.

The prospect of combining stellar ages, kinematics and chemical abundance patterns is now a reality \citep[e.g.,][]{Montalban:2021,Verma:2021}. The goal of this paper is to combine stellar ages, chemical abundances and kinematics to further explore the formation and evolution of our Galaxy and to better understand the limitations and prospects of this approach. In Section~\ref{sec:obs} we describe sample selection, observations and data reduction. Section~\ref{sec:analysis} outlines the methods employed in the analysis, while in Section~\ref{sec:resultages} possible sources of age systematics are discussed. Results are presented in Section~\ref{sec:resultsanddiscussion}, and Section~\ref{sec:conclusions} contains our final remarks.

\section{Observations and data reduction}
\label{sec:obs}

\begin{table*}
	\centering
	\caption{Program stars, IDs from \citet{Epstein:2014}, as well as IDs adopted here for SAGA \citep{Casagrande:2014aa} stars, their adopted average seismic parameters $\Delta \nu_{\mathrm{obs}}$ and $\nu_{\mathrm{max}}$, and their sources: (A) \citet{Yu:2018}, (B) \citet{Pinsonneault:2018}. Period spacings $\Pi_{s}$ from \citet{Ting:2018,Ting:2018erratum} and \citet[][for KIC\,4671239]{Mosser:2018}. Both Gaia DR2 and EDR3 parallaxes \citep{GaiaDR2:2018,GaiaEDR3:2021} shown here are corrected according to \citet{Zinn2019} and \citet{Lindegren:2021}, respectively.}
	\label{tab:tab1}
	\begin{tabular}{rrrrrrrrrr}
		\hline
		Star & ID & RA & DEC & $\Delta \nu_{\mathrm{obs}}$ & $\nu_{\mathrm{max}}$ & Src & $\Pi_{s}$ & $\varpi_{\mathrm{DR2}}$ & $\varpi_{\mathrm{EDR3}}$ \\
		& & (J2000) & (J2000) & $\mu$Hz & $\mu$Hz & & s & mas & mas \\
        \hline
               KIC\,4345370 &  H & 18 59 35.1716 & +39 26 48.895 &  4.050 $\pm$ 0.012 &   32.66 $\pm$   0.29 & A &    76.08 & 0.7693 $\pm$ 0.0235 & 0.6852 $\pm$ 0.0105 \\
               KIC\,4671239 & S1 & 19 43 40.9842 & +39 42 44.763 &  9.782 $\pm$ 0.027 &   98.94 $\pm$   0.44 & A &    66.60 & 0.4489 $\pm$ 0.0227 & 0.4136 $\pm$ 0.0125 \\
               KIC\,6279038 & S2 & 19 19 18.1907 & +41 39 57.419 &  1.154 $\pm$ 0.021 &    5.71 $\pm$   0.03 & B & $\cdots$ & 0.1928 $\pm$ 0.0271 & 0.1898 $\pm$ 0.0099 \\
               KIC\,7191496 &  A & 19 16 07.6460 & +42 46 32.003 &  2.455 $\pm$ 0.021 &   16.23 $\pm$   0.24 & A &    58.76 & 0.4718 $\pm$ 0.0190 & 0.3980 $\pm$ 0.0113 \\
               KIC\,8017159 &  C & 19 08 41.6306 & +43 52 10.320 &  0.690 $\pm$ 0.017 &    3.07 $\pm$   0.02 & B & $\cdots$ & 0.3748 $\pm$ 0.0258 & 0.3217 $\pm$ 0.0093 \\
              KIC\,11563791 &  D & 19 36 52.2056 & +49 35 59.379 &  5.005 $\pm$ 0.019 &   43.03 $\pm$   0.51 & A &    70.09 & 1.0271 $\pm$ 0.0251 & 0.9603 $\pm$ 0.0102 \\
              KIC\,12017985 &  B & 19 37 12.0086 & +50 24 55.593 &  2.620 $\pm$ 0.018 &   18.24 $\pm$   0.29 & A &    63.72 & 0.8540 $\pm$ 0.0269 & 0.8413 $\pm$ 0.0332 \\

	    \hline
	\end{tabular}
\end{table*}

\begin{figure}
	\includegraphics[width=\columnwidth]{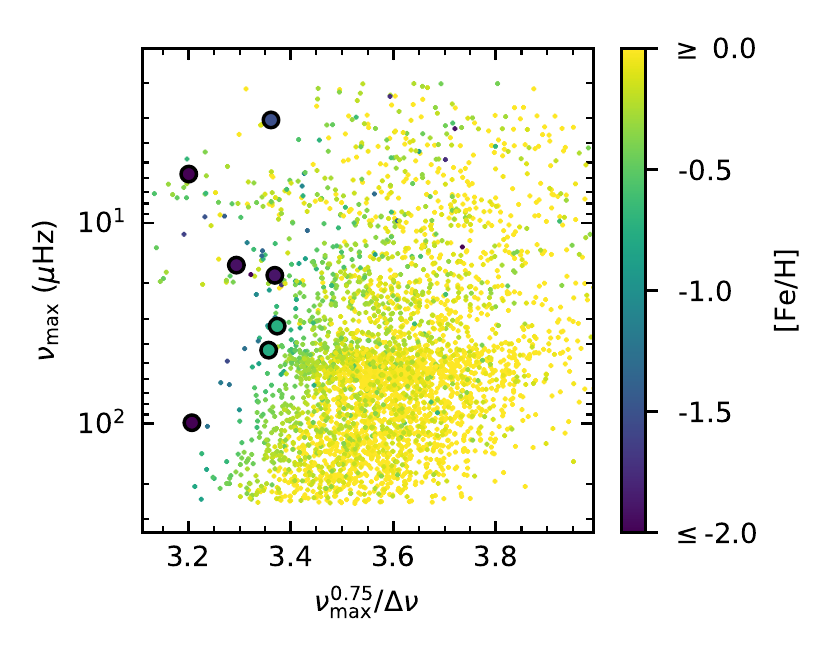}
    \caption{Asteroseismic diagram showing program stars (circles) compared to the subsample of RGB stars from APOKASC-2 \citep[][points]{Pinsonneault:2018}.
    }
    \label{fig:sample}
\end{figure}

The sample used in this work is composed of seven red giant stars in the \emph{Kepler} field whose previous asteroseismic estimates of masses/ages do not appear consistent with their chemical abundance ratios. Five of them are from the first APOKASC catalog \citep{Pinsonneault:2014aa}, and were originally studied by \citet{Epstein:2014}, who derived asteroseismic masses larger than expected by the authors for metal-poor, halo stars. The other two are from SAGA \citep{Casagrande:2014aa}, whose results also suggest young ages for metal-poor stars (e.g., stellar mass of 1.07 M$_{\odot}$ and [Fe/H] = -2.44 in case of KIC\,4671239). Table~\ref{tab:tab1} presents their values for $\Delta \nu$ and $\nu_{\mathrm{max}}$ adopted from the literature. Their position along the RGB can be inspected in Fig.~\ref{fig:sample} through their respective $\nu_{\mathrm{max}}$ values.

These stars were observed with HIRES at W. M. Keck Obserbvatory \citep{Vogt:1994} with the goal of deriving a complete chemical inventory to provide nucleosynthetic checks on ages \citep[program ID: Z079Hr,][]{2016koa..prop..412C}. The raw data were reduced using the Mauna Kea Echelle Extraction (\texttt{MAKEE}) data reduction pipeline, including bias subtraction, flat-fielding, optimal extraction of spectra and wavelength calibration (heliocentric velocity corrected). The individual reduced spectra were radial-velocity corrected, co-added and normalised using \texttt{IRAF}\footnote{\texttt{IRAF} is distributed by the National Optical Astronomy Observatory, which is operated by the Association of Universities for Research in Astronomy, Inc., under cooperative agreement with the National Science Foundation.}. The combined signal-to-noise-ratio (S/N) of the spectra ranges from 60 to 170 at $\lambda \approx$~590~nm, while the nominal spectral resolution is $\lambda$/$\Delta \lambda$ = 47,700. Wavelength coverage is 420-850~nm, with a gap in the 530-550~nm region.

\section{Analysis}
\label{sec:analysis}

The stellar parameters in this work were derived combining different methods run in several iterations: the classical spectroscopic method (see, e.g., \citet[][]{Jofre:2019} for a comprehensive review), the InfraRed Flux Method \citep[IRFM, ][]{Blackwell:1977aa} and asteroseismic grid-based modelling \citep[GBM,][]{Stello:2009}. Spectroscopy allows us to find the detailed chemical composition of a star, while the knowledge of $\Delta \nu$ and $\nu_{\mathrm{max}}$ enables  the possibility of deriving masses with good precision, which then can be converted into ages for an assumed set of isochrones.

\subsection{Fundamental parameters and chemical abundances}

\begin{table*}
	\centering
	\caption{Adopted line list and equivalent widths measured in this work. Lines fitted through spectral synthesis are indicated in the fifth column. Log~gf values are from the NIST database. The complete version of this table is available in the supplementary material.}
	\label{tab:ew}
	\begin{tabular}{rrrrrrrrrrr}
		\hline
		Wavelength & Species &  LEP & log gf & KIC4345370 & KIC4671239 & KIC6279038 & KIC7191496 & KIC8017159 & KIC11563791 & ... \\
               \AA &   Z.ion &   eV &    dex &       m\AA &       m\AA &       m\AA &       m\AA &       m\AA &        m\AA & ... \\
        \hline
           6707.76 &     3.0 & 0.00 &  -0.002 & Synthesis & & & & & & \\
           6707.91 &     3.0 & 0.00 &  -0.303 & Synthesis & & & & & & \\
           6300.30 &     8.0 & 0.00 &  -9.720 & Synthesis & & & & & & \\
           6363.78 &     8.0 & 0.02 & -10.190 & Synthesis & & & & & & \\
           5682.64 &    11.0 & 2.10 &  -0.706 &       71.9 &     ..... &      ..... &      ..... &      ..... &        48.8 & ... \\
           5688.20 &    11.0 & 2.10 &  -0.452 &       90.4 &     ..... &      ..... &      ..... &      ..... &        59.8 & ... \\
           6154.23 &    11.0 & 2.10 &  -1.547 &       24.4 &     ..... &      ..... &      ..... &      ..... &       ..... & ... \\
           6160.75 &    11.0 & 2.10 &  -1.246 &       35.2 &     ..... &      ..... &      ..... &      ..... &        16.3 & ... \\
               ... &     ... &  ... &     ... &        ... &       ... &        ... &        ... &        ... &         ... & ... \\
	    \hline
	\end{tabular}
\end{table*}

For the spectroscopic component of the analysis we have employed solar-scaled 1D LTE Kurucz models interpolated from a grid \citep{Castelli:2003aa}. Initially, the equivalent widths (EW) of \ion{Fe}{I} and \ion{Fe}{II} were measured though Gaussian fitting in selected weak lines compliled from the literature and using log~gf values from NIST\footnote{\url{https://physics.nist.gov/PhysRefData/ASD/lines_form.html}} \citep[][see Table~\ref{tab:ew} for the line list]{Cayrel:2004aa,Alves-Brito:2010aa,Melendez:2012,Barbuy:2013,Ishigaki:2013,Yong:2014} and used as input to find the four atmospheric model parameters with the 2017 version of \texttt{MOOG} \citep{Sneden:1973phd}. Effective temperatures (T$_{\mathrm{eff}}$) and surface gravity values (in terms of log g) were obtained by imposing excitation and ionisation balance, respectively, for each star. The value for the microturbulent velocity ($v_t$) is found eliminating any correlation between \ion{Fe}{I} line strengths -- given by the logarithm of its EW normalized by its wavelength -- and the respective \ion{Fe}{I} abundances as calculated by \texttt{MOOG}.

The adotped models have solar-scaled composition. To account for $\alpha$-enhancement, the metallicity, in terms of [M/H], was estimated using the formula from \citet[][]{Salaris:1993}:

\begin{equation}\label{eq:Salaris}
    \mathrm{[M/H] = [Fe/H] + log_{10}(0.638 \times 10^{[\alpha/Fe]} + 0.362)}
\end{equation}

\noindent where $\alpha$ is the median of (Mg, Si, Ca). \texttt{MOOG} calculates the abundances from the EW of the individual species, allowing us to estimate the output [M/H]$_{\mathrm{(Fe, \alpha)}}$ with Eq.~\ref{eq:Salaris}. The metallicity is then found iteratively when an atmospheric model yields a [M/H]$_{\mathrm{(Fe, \alpha)}}$ that is equal to the input [M/H]$_{\mathrm{MODEL}}$. The spectroscopic atmospheric parameters are found when they generate an atmospheric model that satisfies \emph{simultaneously} all the conditions described above -- (i) excitation balance, (ii) ionization balance, (iii) absence of correlation between line strength and line abundance and (iv) agreement between input and output [M/H].\footnote{Atmospheric parameters convergence was achieved using the \texttt{MOOG} wrapper \texttt{Xiru}, written in Python, available at \url{https://github.com/arthur-puls/xiru}.}

\begin{figure}
	\includegraphics[width=\columnwidth]{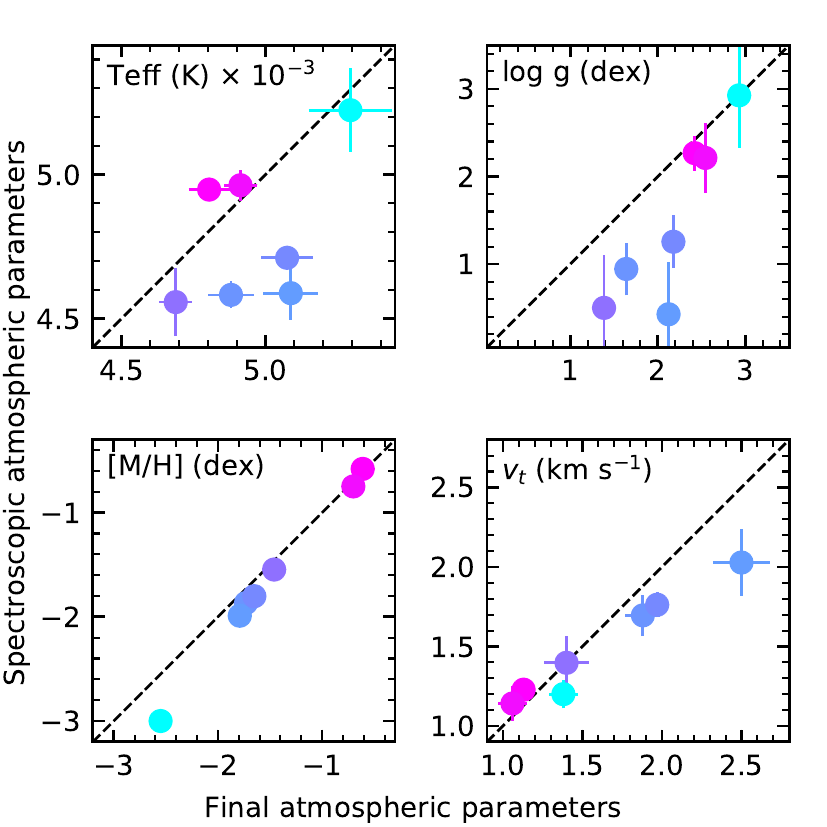}
    \caption{Pure spectroscopic parameters as function of respective adopted atmospheric parameters shown in Table~\ref{tab:atmparam}. Points are color-coded by adopted [M/H].
    Dashed lines represent the identity function, while dotted lines show typical uncertainty intervals for the pure spectroscopic set.
    }
    \label{fig:atmcomp}
\end{figure}

For each star, the value of [Fe/H] used to estimate [M/H] is the median of the abundances taken from its \ion{Fe}{II} lines. The adoption of \ion{Fe}{II} lines for [M/H] has been made due to the large metallicity range of the program stars and the lower sensitivity of \ion{Fe}{II} to NLTE effects. However, the small number of measurable \ion{Fe}{II} lines (see Table~\ref{tab:ew}) prevents a reliable linear fit to be made between line strengths and abundances and thus we have had to rely on the larger number of \ion{Fe}{I} lines for this task. It is known that the $\Delta \mathrm{[Fe/H] = [Fe/H]_{NLTE} - [Fe/H]_{LTE}}$ is different for each of the individual \ion{Fe}{I} lines and this can impact the estimation of $v_t$ \citep[e.g.,][]{Lind:2012aa,Amarsi:2016aa}. \citeauthor{Lind:2012aa} estimated the change in $v_t$ due to NLTE effects, and from inspection of their fig.~7 we can assume that these effects are very mild in the parameter space of our sample, $|\Delta v_t| \rightarrow 0$ km s$^{-1}$. The spectrum of the most metal-poor target, KIC\,4671239, shows no \ion{Fe}{II} features. In this case, derivation of atmospheric parameters through full ionisation/excitation balance is not possible. 'Spectroscopic' atmospheric parameters for KIC\,4671239 were derived keeping [M/H] fixed at -3~dex and adopting log~g from Eq.~\ref{eq:logg_sc}, and their respective uncertainties (shown in Fig.~\ref{fig:atmcomp}) are conservative estimates. We tested two representative stars, KIC\,8017159 and KIC\,12017985, to check if the adoption of alpha-enhanced models instead of solar-scaled models with Eq.~\ref{eq:Salaris} would result in qualitatively distinct results. In both cases, [X/H] have consistently shifted by $\delta$[X/H] $\approx$ $-$0.08~dex for the species measured with the EW method, with scatter in $\delta$[X/H] being lower than 0.02~dex. Hence, [X/Fe] seems to be unaffected by the choice of models. These two stars were chosen because of their distinct [Mg/Fe], the former having [Mg/Fe] close to the solar value and the later being Mg-rich (see Section~\ref{sec:resultsanddiscussion} for the results).

With initial values for metallicity calculated using the classical spectroscopic method, we could then estimate T$_{\mathrm{eff}}$ for each star using an implementation of the IRFM \citep{Casagrande:2010aa,Casagrande:2021}. This process was divided in two steps. For the first one, input photometry was taken from published JHKs 2MASS \citep{Cutri:2003} and Gaia DR2 magnitudes \citep{GaiaDR2:2018}. Reddening was estimated using the 2019 version of Bayestar \citep{Green:2019}, interpolated over distances published by \citet[][]{BailerJones:2018}. The IRFM generates a grid of T$_{\mathrm{eff}}$ as function of metallicity and log g and the resulting temperatures were interpolated from this grid using the values calculated with the classical spectroscopic method described above.

Before proceeding to the second step of the IRFM we made a run of the first iteration of GBM employing the version 0.27 of \texttt{BASTA} \citep{SilvaAguirre:2015,SilvaAguirre:2017,Aguirre:2021}. We used \texttt{BASTA} to fit a set of observables into a grid of BaSTI isochrones \citep{Hidalgo:2018} to estimate the fundamental parameters of a star using bayesian inference. The following parameters were used as global inputs for the BaSTI isochrones: overshooting (0.2), diffusion (0.0), $\eta$ mass loss (0.3) and no $\alpha$-enhancement. The mass loss parameter is still very uncertain \citep[e.g.,][]{Miglio:2012,Tailo:2020} and, while the value $\eta$=0.3 was adopted, we also tested the situation when mass-loss is neglected (see Section~\ref{sec:resultages}). Meanwhile, $\alpha$-enhancement was defined as zero because we are scaling metallicity to the solar mixture with Eq.~\ref{eq:Salaris}.

The model-based correction from \citet{Serenelli:2017} has been applied for $\Delta\nu$, rescaling this observable though a correction factor $f_{\Delta\nu}$:

\begin{equation}\label{eq:dnucor}
    \Delta\nu_{\mathrm{m}} = \frac{\Delta\nu}{f_{\Delta\nu}}
\end{equation}

\noindent where $\Delta\nu$ represents the value of the large frequency separation for a given star listed in Table~\ref{tab:tab1}, while $\Delta\nu_{\mathrm{m}}$ is the value used for BaSTI grid fitting. From Eqs.~\ref{eq:mass_sc} and \ref{eq:dnucor}, the ratio between corrected and uncorrected stellar masses is expected to be $f_{\Delta\nu}^{4}$ unless further corrections are applied. Typical values of $f_{\Delta\nu}$ range from 0.95 to 1, i.e., the corrected mass can be up to 20\% lower than the uncorrected one.

Regarding the individual observables, two sets were chosen: s$_p$ = (T$_{\mathrm{eff}}$, [M/H], $\Delta\nu$, $\nu_{\mathrm{max}}$, evolutionary phase, 2MASS $K_s$, parallax) and another set without parallax, hereafter called s$_n$. To inform the prior on evolutionary stage the values of period spacing shown in Table~\ref{tab:tab1} were adoped. All targets in this work with period spacing measurements $\Pi_{s}$  have $\Delta \nu <$~10~$\mu$Hz and $\Pi_{s} <$~100~s. All of them (5 out of 7) can be assumed as being on the RGB \citep[see, e.g, fig. 1 from][]{Mosser:2014}. For the remaining stars, KIC\,6279038 and KIC\,8017159, the evolutionary stage recommended by \citet{Yu:2018} was adoped (RGB) for the former, while the later had no evolutionary stage prior defined. Flat priors were adopted for the other observables. The output from GBM yielded current and initial masses, radii, luminosities, ages, E($B-V$), distances and log g, as well as fitted values for the input observables. These log g values, along with the E($B-V$) calculated with \texttt{BASTA}, were then used as input in the second iteration of the IRFM.

\begin{table}
	\centering
	\caption{The stellar parameters adopted to generate the atmospheric models for the spectroscopic analysis.}
	\label{tab:atmparam}
	\begin{tabular}{lrrrr}
		\hline
		Star & T$_{\mathrm{eff}}$ & log g & [M/H] & $v_t$ \\
             & K & dex & dex & km s$^{-1}$ \\
        \hline
             KIC\,4345370 & 4804 & 2.421 & -0.61 & 1.13 \\
             KIC\,4671239 & 5295 & 2.929 & -2.55 & 1.38 \\
             KIC\,6279038 & 4880 & 1.640 & -1.73 & 1.88 \\
             KIC\,7191496 & 5088 & 2.123 & -1.79 & 2.50 \\
             KIC\,8017159 & 4688 & 1.386 & -1.46 & 1.40 \\
            KIC\,11563791 & 4913 & 2.543 & -0.70 & 1.06 \\
            KIC\,12017985 & 5075 & 2.178 & -1.65 & 1.97 \\
	    \hline
	\end{tabular}
\end{table}

The second IRFM iteration yielded the T$_{\mathrm{eff}}$ values adopted in this work. The median difference from the first to the second iteration is 17~K. The final T$_{\mathrm{eff}}$ values replaced those from the first IRFM iteration in a new run of \texttt{BASTA} that yielded the values adopted in this work for the output parameters mass -- both the present-day asteroseismically inferred one, and initial one before any mass-loss took place -- radius, luminosity, age, E($B-V$), distance and log g. The surface gravity and T$_{\mathrm{eff}}$ values derived in this last \texttt{BASTA}/IRFM run were employed to calculate the final [M/H] and $v_t$. Both [M/H] and $v_t$ were calculated in the same way as done in the classical spectroscopic approach, but this time with T$_{\mathrm{eff}}$ and log~g kept fixed to IRFM/\texttt{BASTA} values when interpolating the Kurucz models. Table~\ref{tab:atmparam} shows the final stellar parameters adopted in this work. In Fig.~\ref{fig:atmcomp} the atmospheric parameters derived using the classical spectroscopic method are compared with the adopted parameters. The metal-poor stars ([M/H] $<$ -1) present the largest deviations between the purely spectroscopic and hybrid \footnote{Spectroscopic metallicity and microturbulence, with IRFM T$_{\mathrm{eff}}$ and asteroseismic log~g.} sets of atmospheric parameters.

Using the atmospheric model interpolated from the adopted T$_{\mathrm{eff}}$, log g, [M/H], and $v_t$, the abundances for each species could be calculated, either by direct curve of growth analysis using the \texttt{MOOG} driver \emph{abfind} or by fitting a synthetic spectrum on the observed one. Species whose lines are subject to hyperfine and/or isotopic splitting, or are known to have non-negligible blends, were measured by spectral synthesis (Li, O, odd-Z Fe-peak, and n-capture elements) by fitting synthetic spectra to the observed ones. The remaining species had their abundances calculated through curve of growth analysis.

\subsubsection{Uncertainties}
\label{sec:unc}

Uncertainties for the atmospheric parameters were estimated as follows: for T$_{\mathrm{eff}}$, a MonteCarlo (MC) was run in the IRFM using uncertainties in the adopted [Fe/H], log(g), reddening and photometry. The surface gravity uncertainties come from the 16- and 84- percentiles of the posterior calculated by \texttt{BASTA}. For [M/H] and $v_t$ uncertainties, we adapted the method outlined by \citet{Epstein:2010aa}.

Summarising their approach, the classical spectroscopic method yields a vector of atmospheric model parameters:

\begin{equation}\label{eq:atmparvec}
    \mathbf{m} = (\mathrm{T_{eff}}, \mathrm{log\:g}, \mathrm{[M/H]}, v_t),
\end{equation}

\noindent which is found when all components of the observables vector $\mathbf{o}$ converge to a zero vector. The components of $\mathbf{o}$ are:

\begin{itemize}
    \item $o_1$: slope of the linear fit from line-by-line \ion{Fe}{I} abundances measured against their respective lower excitation potentials;
    \item $o_2$: difference, in dex, between the medians of \ion{Fe}{I} and \ion{Fe}{II} abundances;
    \item $o_3$: difference, in dex, between [M/H]$_{\mathrm{(Fe, \alpha)}}$ and [M/H]$_{\mathrm{MODEL}}$, and
    \item $o_4$: slope of the linear fit from line-by-line \ion{Fe}{I} abundances measured against the logarithm of their respective equivalent widths normalized by their wavelengths.
\end{itemize}

To propagate the uncertainties from $\mathbf{o}$ to $\mathbf{m}$, a matrix $\mathbf{B}$ is built making $B_{ij}$ = $\partial o_i/\partial m_j$. The values for $B_{ij}$ adopted here were calculated using the spectral atlas from Arcturus \citep{Hinkle:2000} and are slightly different from those shown in table 2 from \citet{Epstein:2010aa}. The propagation of uncertainties is then done by:

\begin{equation}\label{eq:propag}
    \mathbf{\sigma^{m}} = \mathbf{B^{-1}}\mathbf{\sigma^{o}}
\end{equation}

\noindent where $\mathbf{\sigma^{m}}$ and $\mathbf{\sigma^{o}}$ are the vectors containing the respective uncertainties of $\mathbf{m}$ and $\mathbf{o}$. However, the parameters $m_1$ and $m_2$, T$_{\mathrm{eff}}$ and log g, respectively, were not calculated by the classical spectroscopic method and thus the values of $\sigma^{o}_1$ and $\sigma^{o}_2$ were estimated using the uncertainties for T$_{\mathrm{eff}}$ and log g calculated with the IRFM and \texttt{BASTA}:

\begin{equation}\label{eq:subs}
    \sigma^{o}_i \simeq B_{ii} \sigma^{e}_{i}
\end{equation}

\noindent where $\sigma^{e}_{i}$ are the uncertainties for T$_{\mathrm{eff}}$ and log g calculated with the other methods.

The value of $\sigma^{o}_3$ is calculated by perturbing the line-by-line Fe measurements: a vector $\mathbf{s}$ whose size is the number of Fe lines measured in a given star -- whose measurements form the vector $\mathbf{s}^{\mathrm{Fe}}$ -- is sampled from a normal distribution with the star's A(Fe) as mean and the standard deviation of the line-by-line Fe measurements as its scale. Then $\mathbf{s}^{\mathrm{Fe}}$ is subtracted from $\mathbf{s}$ and the mean $\xi$ of the components of the resulting vector is calculated. The process is repeated 1000 times, and, finally, $\sigma^{o}_3$ is taken as the mean of all 1000 $\xi$ values. The value of $\sigma^{o}_4$ is the uncertainty of the slope of the linear fit used to determine $o_4$.

After calculating $\mathbf{\sigma^{m}}$ with Eq.~\ref{eq:propag}, the values of $\sigma^{m}_{1}$ and $\sigma^{m}_{2}$ are replaced with the uncertainties for T$_{\mathrm{eff}}$ and log g previously calculated. The adopted uncertainties for the atmospheric parameters are shown in Table~\ref{tab:apdx_atmunc}.

The uncertainties for the individual abundances A(X) are estimated by taking the sensitivities of A(X) to 1-$\sigma$ variation of the atmospheric parameters. Those sensitivities are added in quadrature along with the mean absolute deviation of the line-by-line measurements of the element X, giving the resulting uncertainties $\sigma_{A(X)}$ shown in Table~\ref{tab:apdx_axunc}.

\subsection{Dynamics}

The dynamical properties of the stars were determined following the procedure described in \citet{Monty:2020}. Briefly, we adopted the potential characterised in \citet{McMillan:2017} and implemented in the galactic dynamics Python package, \texttt{GALPY} as \texttt{McMillan17} \citep{Bovy:2015}. The characteristics of the potential components, solar galactocentric distance ($R_{0}=8.21$~kpc), and circular velocity ($v_{0}=233.1$~km/s) given in Table 3 of \citeauthor{McMillan:2017} were all left unchanged. The solar position was set to (X, Y, Z) = (0, 0, 20.8~pc) as given by \citet{Bennett:2019}. We used Gaia EDR3 data for positions, parallaxes \citep[corrected according to][]{Lindegren:2021} and proper motions \citep[][]{GaiaEDR3:2021}. Radial velocities were taken from Gaia DR2 \citep[][]{GaiaDR2:2018} for the six objects with radial velocity measurements available in the catalogue. KIC\,4671239 does not have radial velocity information in Gaia, thus the value of $-$187.5 $\pm$ 2.0 km s$^{-1}$ for heliocentric radial velocity measured in our observed spectrum was adopted. To explore the errors associated with each dynamical property, we constructed the covariance matrix associated with the Gaia EDR3 parameters for each star. A symmetric error distribution was assumed for the radial velocity values. We performed 1000 MC realisations for each star to sample the error distributions. Stellar orbits were integrated for each realisation, integrating forwards and backwards 5-7 Gyr for a total integration time of 10-14 Gyr. 

As in \citet{Monty:2020}, we recover the actions ($J_{\phi}$, $J_{R}$, $J_{z}$), orbital period ($T$), eccentricity ($e$), and maximum height from the galactic plane ($Z_{\mathrm{max}}$) for each star using an implementation of the St{\"a}ckel Fudge in \texttt{GALPY} \citep{Binney:2012,Mackereth:2018}. The three dimensional velocities $U$, $V$ and $W$ were also determined after accounting for the solar peculiar velocity as determined by \citet{Schonrich:2010} ($U, V, V$) = ($11.1, 12.24, 7.25$~km/s). Finally, to compare with the inclination values of \cite{Yuan:2020}, we estimate the orbital inclination as the orbit-averaged angle between the total angular momentum $|L_{\mathrm{tot}}|$ and the $z$-component of the angular momentum $L_{z}$.

\section{Ages from grid-based modelling}
\label{sec:resultages}

\begin{table}
	\centering
	\caption{Ages, masses and radii derived with the s$_n$ (T$_{\mathrm{eff}}$, [M/H], $\Delta\nu$, $\nu_{\mathrm{max}}$, evolutionary phase, 2MASS $K_s$) set, as well as adopted [Fe/H].}
	\label{tab:epstein}
	\begin{tabular}{rrrrr}
		\hline
		KIC & Age & Mass & Radius & $[$Fe/H$]$ \\
            & (Gyr) & (M$_{\odot}$) & (R$_{\odot}$) & dex \\
        \hline
             4345370 &  9.2$^{+1.1}_{-0.9}$ & 0.95$^{+0.02}_{-0.03}$ &  9.9$^{+0.1}_{-0.1}$ & -0.75 \\
             4671239 &  5.4$^{+0.4}_{-0.3}$ & 1.01$^{+0.02}_{-0.02}$ & 5.70$^{+0.05}_{-0.05}$ & -2.63 \\
             6279038 & 13.8$^{+3.4}_{-3.1}$ & 0.76$^{+0.06}_{-0.05}$ & 21.3$^{+0.8}_{-0.6}$   & -2.11 \\
             7191496 &  8.7$^{+1.6}_{-2.1}$ & 0.88$^{+0.08}_{-0.04}$ & 13.5$^{+0.5}_{-0.3}$ & -1.92 \\
             8017159 & 10.1$^{+4.7}_{-3.9}$ & 0.83$^{+0.12}_{-0.09}$ & 30.7$^{+2.1}_{-1.6}$ & -1.52 \\
            11563791 &  8.8$^{+1.5}_{-1.3}$ & 0.94$^{+0.04}_{-0.04}$ &  8.6$^{+0.1}_{-0.1}$ & -0.78 \\
            12017985 &  5.5$^{+1.3}_{-1.1}$ & 1.01$^{+0.07}_{-0.06}$ & 13.6$^{+0.3}_{-0.3}$ & -1.89 \\
        \hline
	\end{tabular}
\end{table}

\begin{figure*}
	\includegraphics{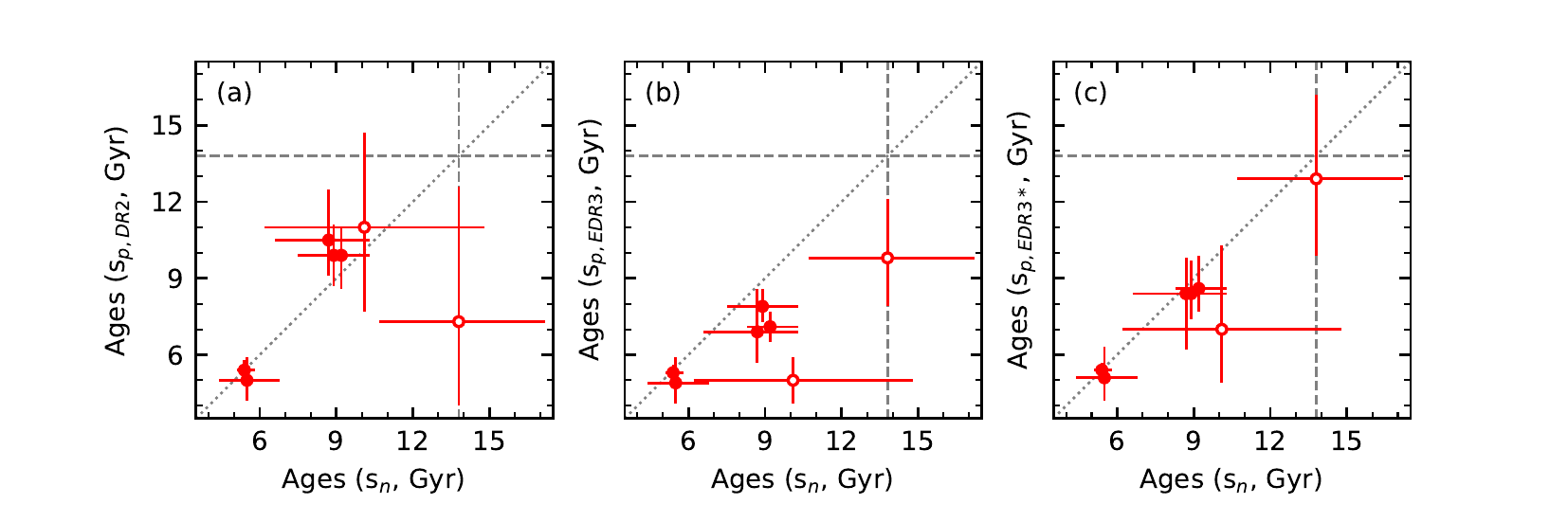}
    \caption{Comparison between ages derived with s$_n$ (T$_{\mathrm{eff}}$, [M/H], $\Delta\nu$, $\nu_{\mathrm{max}}$, evolutionary phase, 2MASS $K_s$), and ages calculated with s$_p$ (i.e., s$_n$ plus parallax) using different parallax inputs from Gaia.
    \emph{(a)}: s$_p$ adopts DR2 parallaxes;
    \emph{(b)}: s$_p$ adopts EDR3 parallaxes;
    \emph{(c)}: same as (b), but inflating the parallax uncertainties by 0.02 mas.
    Dashed lines represent the age of the Universe \citep{Planck:2020}, while dotted lines show the identity function.
    Filled circles have asteroseismic parameters compiled from \citet{Yu:2018}, while open circles represent stars with parameters from \citet{Pinsonneault:2018}.
    }
    \label{fig:ages_and_plx}
\end{figure*}

As explained in the Section~\ref{sec:analysis}, in this work we tested two sets of observables, one -- s$_p$ -- that does include parallaxes and a second one -- s$_n$ -- that does not. In s$_p$ we have made three runs using Gaia parallaxes: (a) employing values published by Gaia DR2 \citep{GaiaDR2:2018} and corrected according to \citet{Zinn2019}; (b) using parallaxes from Gaia EDR3 \citep{GaiaEDR3:2021} and corrections from \citet{Lindegren:2021}; and, finally, (c) also using EDR3 parallaxes, but this time increasing the published uncertainties by 0.02 mas (in order to decrease the weight of parallax in the solution). The effect of these changes can be seen in Fig.~\ref{fig:ages_and_plx}, where the ages of these three different s$_p$ runs are plotted as function of s$_n$ ages.

Fig.~\ref{fig:ages_and_plx}(a) shows agreement between s$_p$ and the pure asteroseismic solution s$_n$ within the error bars, except for KIC\,6279038 -- the oldest star in the s$_n$ set. The overall agreement between s$_n$ and s$_p$ seen in Fig.~\ref{fig:ages_and_plx}(a) is expected since the parallax corrections employed in this s$_p$ set have been calibrated with asteroseismology \citep{Zinn2019}.

When Gaia EDR3 parallaxes are employed, as shown in Fig.~\ref{fig:ages_and_plx}(b), s$_p$ error bars decrease, but the 1-$\sigma$ agreement disappear for several stars. In particular, KIC\,8017159 -- the second-oldest in the s$_n$ set -- now displays a very small uncertainty in s$_p$ age, but with the median age now being similar to the youngest s$_n$ object at $\sim$~5~Gyr. Also, using parallaxes now yields lower ages to all targets, while the opposite happens to most of our stars when DR2 parallaxes are employed.

In order to check if the results found with Gaia EDR3 parallaxes might be affected by some systematics we did another s$_p$ run with EDR3 error bars increased by 0.02 mas -- making them slightly larger than those from DR2, whose results are shown in Fig.~\ref{fig:ages_and_plx}(c). With the slightly larger error bars resulting from the increase in the uncertainty of the parallax, the relationship between s$_p$ and s$_n$ is closer to 1:1, including for KIC\,6279038. However, KIC\,8017159 keeps being younger, albeit with the error bar now touching the identity function. The overall result suggests that Gaia EDR3 parallax uncertainties might be underestimated if the pure seismic solutions are reliable, and employing them without any corrections might result in trading accuracy for precision in several cases, at least in the lower (M $\lesssim$ 1.0~M$_{\odot}$) mass regime. These results converge to agreement with findings both from the Gaia team and independent studies that the published parallax uncertainties from Gaia EDR3 might be underestimated at least for some regions of the parameter space \citep{El-Badry:2021,Fabricius:2021,Zinn:2021aj}.

Regarding KIC\,8017159, the DR2-EDR3 parallax difference is 0.05 mas while the uncertainty in EDR3 is $\sigma_{\varpi} \approx$ 0.01 mas, i.e., the difference is of the order of 5-$\sigma$. Its EDR3 RUWE is 0.88, while other astrometric quality flags suggested by \citet{Fabricius:2021} such as \texttt{ipd\_harmonic\_amplitude}, \texttt{ipd\_muli\_peak}, \texttt{phot\_bp\_rp\_excess\_factor}, and the excess noise of the source \texttt{astrometric\_excess\_noise} also display good values, thus the astrometric solution is expected to be good.

However, other program stars have had even larger variations between DR2 and EDR3 parallaxes, as seen in Table~\ref{tab:tab1}, but their GBM solutions were not as sensitive to parallax as in the case of KIC\,8017159. According to the results from our GBM, this object and KIC\,6279038 are upper-RGB stars -- i.e. RGB stars whose $\nu_{\mathrm{max}}$ is lower than 10~$\mu$Hz. The result for the evolutionary stage is always RGB for all stars in this work regardless of the chosen prior (RGB or none). As shown in Fig.~\ref{fig:ages_and_plx}, these objects on the upper-RGB also have the worst precision in the modelled ages, regardless of the chosen input set. \citet{Zinn:2019apj885} have found a departure from the scaling relation for stellar radii on the upper-RGB, suggesting that we might not understand very well the relationship between stellar pulsations and fundamental parameters in this region of the HR diagram. However, it must be noted that this departure has been identified for radii $>$ 30~R$_{\odot}$, while our program stars have R $\lesssim$ 30~R$_{\odot}$.

Given the large change in Gaia parallaxes from DR2 to EDR3 in some of our stars and the resulting differences in modelled ages that have arisen from that, adopting the ages derived from the pure seismic solution (s$_n$) seems to be the safest option in this study. The choice of one set over the other has negligible impact on the atmospheric parameters used in the spectroscopic analysis -- the change in log~g is of the order of 0.01~dex, and the resulting impact on abundances is negligible. Nevertheless, the ages estimated for stars on the upper RGB must be interpreted with caution.

In order to do a sanity check, we also compare the results derived with the s$_n$ input set with results taken from \texttt{PARAM} \citep{daSilva:2006,Rodrigues:2014}\footnote{Using the web form available at \url{http://stev.oapd.inaf.it/cgi-bin/param}.}, using the same observational inputs and this time employing the models from \citet{Rodrigues:2017}. Modelling with \texttt{PARAM} yields larger stellar masses for three objects, although both \texttt{PARAM} and \texttt{BASTA} error bars overlap the identity function with the exception of KIC\,11563791, whose \texttt{PARAM} age is 7.0$^{+1.1}_{-0.9}$~Gyr and \texttt{PARAM} mass is 0.98$^{+0.04}_{-0.03}$~M$_{\odot}$. These slightly larger masses correspond to (slightly) lower ages derived by \texttt{PARAM}. One notable object in this comparison is KIC\,6279038, which shows a negligible difference in stellar mass and an 1~Gyr difference in (median) age.

In their study of asteroseismology of eclipsing binaries, \citet{Brogaard:2018} noted that the models from \citet{Rodrigues:2017} are possibly too cool in T$_{\mathrm{eff}}$. This could lead to overestimation of stellar mass for a given observed T$_{\mathrm{eff}}$ value. The difference in the masses derived for KIC\,11563791 corresponds to a T$_{\mathrm{eff}}$ shift of -100~K in the scaling relations, assuming the same $f_{\Delta\nu}$ in the shifted and non-shifted cases. Another possible source of the (small) disagreement seen between \texttt{BASTA} and \texttt{PARAM} are differences in each approach of GBM \citep[see, e.g.,][for a comprehensive discussion]{Serenelli:2017}.

In Section~\ref{sec:analysis} it is mentioned that the adopted mass loss prescription is $\eta$ = 0.3. All stars in this study are still ascending the RGB, thus we expect negligible effects because their radii are $\lesssim$~30~R$_{\odot}$. In order to check the effect of different choices of the mass loss parameter we also performed a s$_n$ run, this time using $\eta$ = 0. Intuitively, the exclusion of mass loss should result in older ages, because the current asteroseismic mass is supposed to be the same in both cases, i.e., in the $\eta$ = 0 scenario the initial mass would be lower. This is observed for core He-burning stars, but not on the RGB, where there is no clear conclusion \citep[see][]{Casagrande:2016aa}. When isochrones with and without mass loss are plotted against each other in a Kiel diagram, the model with $\eta$ = 0 is slightly cooler in the upper RGB. Thus, for a fixed set of observables, the $\eta$ = 0 scenario would require a younger isochrone (relative to $\eta$ = 0.3) to fit the input parameters. However, the difference between the isochrones in this evolutionary stage (1.3 $<$ log~g $<$ 2.0) is small enough that the random uncertainties in the input observables likely dominate over different choices of mass loss prescription. Indeed, in our $\eta$ = 0 test, no clear systematic shift has been found in stellar ages, nor large ($\geq$ 1-$\sigma$) deviations in the median ages have been detected.

\section{Results and discussion}
\label{sec:resultsanddiscussion}

\begin{table*}
	\centering
	\caption{Abundance ratios [X/Fe], A(Li), and [Fe/H].}
\label{tab:abundances1}
\begin{tabular}{rrrrrrrrrr}
\hline
KIC & A(Li) &  [O/Fe] & [Na/Fe] & [Mg/Fe] & [Al/Fe] & [Si/Fe] & [Ca/Fe] & [Sc/Fe] & [Ti/Fe] \\
\hline
 4345370 & 0.92 $\pm$ 0.12 & 0.47 $\pm$ 0.07 &  0.02 $\pm$ 0.10 &  0.22 $\pm$ 0.08 &  0.21 $\pm$ 0.08 &  0.27 $\pm$ 0.06 &  0.05 $\pm$ 0.10 &  0.10 $\pm$ 0.07 &  0.02 $\pm$ 0.08 \\
 4671239 &        $\cdots$ &        $\cdots$ &         $\cdots$ &  0.11 $\pm$ 0.18 &         $\cdots$ &         $\cdots$ &  0.14 $\pm$ 0.17 &         $\cdots$ &  0.29 $\pm$ 0.20 \\
 6279038 &        $\cdots$ & 1.11 $\pm$ 0.15 &         $\cdots$ &  0.45 $\pm$ 0.14 &         $\cdots$ &         $\cdots$ &  0.52 $\pm$ 0.14 &  0.14 $\pm$ 0.15 &  0.43 $\pm$ 0.17 \\
 7191496 &        $\cdots$ &        $\cdots$ &         $\cdots$ &         $\cdots$ &         $\cdots$ &         $\cdots$ &  0.15 $\pm$ 0.15 &  0.15 $\pm$ 0.11 &  0.39 $\pm$ 0.17 \\
 8017159 &        $\cdots$ & 0.50 $\pm$ 0.12 &         $\cdots$ &  0.02 $\pm$ 0.11 &         $\cdots$ &  0.08 $\pm$ 0.12 &  0.14 $\pm$ 0.15 & -0.06 $\pm$ 0.11 & -0.14 $\pm$ 0.23 \\
11563791 &        $\cdots$ & 0.50 $\pm$ 0.13 & -0.29 $\pm$ 0.13 &  0.08 $\pm$ 0.13 &         $\cdots$ &  0.26 $\pm$ 0.13 &  0.05 $\pm$ 0.13 & -0.07 $\pm$ 0.13 & -0.12 $\pm$ 0.15 \\
12017985 & 1.23 $\pm$ 0.13 & 0.94 $\pm$ 0.15 & -0.07 $\pm$ 0.15 &  0.36 $\pm$ 0.15 &         $\cdots$ &  0.24 $\pm$ 0.16 &  0.36 $\pm$ 0.15 &  0.07 $\pm$ 0.15 &  0.41 $\pm$ 0.16 \\
\hline
KIC &  [V/Fe] & [Cr/Fe] & [Mn/Fe] & [FeI/H] & [FeII/H] & [Co/Fe] & [Ni/Fe] & [Cu/Fe] & [Zn/Fe] \\
\hline
 4345370 & -0.24 $\pm$ 0.11 & -0.28 $\pm$ 0.07 & -0.64 $\pm$ 0.09 & -0.89 $\pm$ 0.11 & -0.75 $\pm$ 0.07 & -0.10 $\pm$ 0.07 & -0.06 $\pm$ 0.12 &  0.06 $\pm$ 0.08 &  0.71 $\pm$ 0.07 \\
 4671239 &         $\cdots$ &         $\cdots$ &         $\cdots$ & -2.63 $\pm$ 0.20 &         $\cdots$ &         $\cdots$ &         $\cdots$ &         $\cdots$ &         $\cdots$ \\
 6279038 &  0.06 $\pm$ 0.15 &         $\cdots$ & -0.34 $\pm$ 0.15 & -1.94 $\pm$ 0.17 & -2.11 $\pm$ 0.16 &         $\cdots$ &  0.15 $\pm$ 0.14 & -0.38 $\pm$ 0.16 &  0.20 $\pm$ 0.14 \\
 7191496 &         $\cdots$ &         $\cdots$ &         $\cdots$ & -1.98 $\pm$ 0.21 & -1.92 $\pm$ 0.12 &         $\cdots$ &  0.07 $\pm$ 0.12 &         $\cdots$ &  0.10 $\pm$ 0.10 \\
 8017159 &         $\cdots$ &         $\cdots$ & -0.32 $\pm$ 0.12 & -1.56 $\pm$ 0.18 & -1.52 $\pm$ 0.13 &         $\cdots$ & -0.24 $\pm$ 0.11 &         $\cdots$ & -0.03 $\pm$ 0.12 \\
11563791 & -0.29 $\pm$ 0.13 & -0.34 $\pm$ 0.13 & -0.82 $\pm$ 0.13 & -1.03 $\pm$ 0.12 & -0.78 $\pm$ 0.15 &         $\cdots$ & -0.26 $\pm$ 0.14 & -0.36 $\pm$ 0.13 &  0.66 $\pm$ 0.19 \\
12017985 &  0.07 $\pm$ 0.17 &         $\cdots$ & -0.32 $\pm$ 0.15 & -1.79 $\pm$ 0.17 & -1.89 $\pm$ 0.17 &         $\cdots$ &  0.15 $\pm$ 0.15 & -0.35 $\pm$ 0.15 &  0.18 $\pm$ 0.15 \\
\hline
KIC & [Sr/Fe] &  [Y/Fe] & [Zr/Fe] & [Ba/Fe] & [La/Fe] & [Ce/Fe] & [Nd/Fe] & [Sm/Fe] & [Eu/Fe] \\
\hline
 4345370 &  0.25 $\pm$ 0.10 & -0.06 $\pm$ 0.15 & -0.08 $\pm$ 0.14 &  0.14 $\pm$ 0.12 &  0.14 $\pm$ 0.08 &  0.19 $\pm$ 0.15 &  0.35 $\pm$ 0.18 &  0.31 $\pm$ 0.06 &  0.04 $\pm$ 0.06 \\
 4671239 &         $\cdots$ &         $\cdots$ &         $\cdots$ &         $\cdots$ &         $\cdots$ &         $\cdots$ &         $\cdots$ &         $\cdots$ &         $\cdots$ \\
 6279038 &         $\cdots$ & -0.04 $\pm$ 0.14 &         $\cdots$ &  0.07 $\pm$ 0.14 &  0.43 $\pm$ 0.14 &  0.25 $\pm$ 0.15 &  0.54 $\pm$ 0.14 &  0.62 $\pm$ 0.15 &         $\cdots$ \\
 7191496 &         $\cdots$ &  0.06 $\pm$ 0.11 &         $\cdots$ & -0.23 $\pm$ 0.10 &         $\cdots$ &         $\cdots$ &         $\cdots$ &         $\cdots$ &         $\cdots$ \\
 8017159 &         $\cdots$ & -0.21 $\pm$ 0.11 &         $\cdots$ &         $\cdots$ & -0.11 $\pm$ 0.13 & -0.07 $\pm$ 0.12 & -0.13 $\pm$ 0.11 &         $\cdots$ &  0.21 $\pm$ 0.11 \\
11563791 &         $\cdots$ &  0.44 $\pm$ 0.13 &         $\cdots$ &  0.33 $\pm$ 0.15 &  0.28 $\pm$ 0.13 &         $\cdots$ &  0.20 $\pm$ 0.13 &  0.67 $\pm$ 0.14 & -0.15 $\pm$ 0.13 \\
12017985 &         $\cdots$ &  0.08 $\pm$ 0.15 &         $\cdots$ & -0.07 $\pm$ 0.15 &  0.43 $\pm$ 0.15 &  0.34 $\pm$ 0.15 &  0.50 $\pm$ 0.15 &         $\cdots$ &  0.14 $\pm$ 0.16 \\
\hline
\end{tabular}
\end{table*}

\begin{figure}
	\includegraphics[width=\columnwidth]{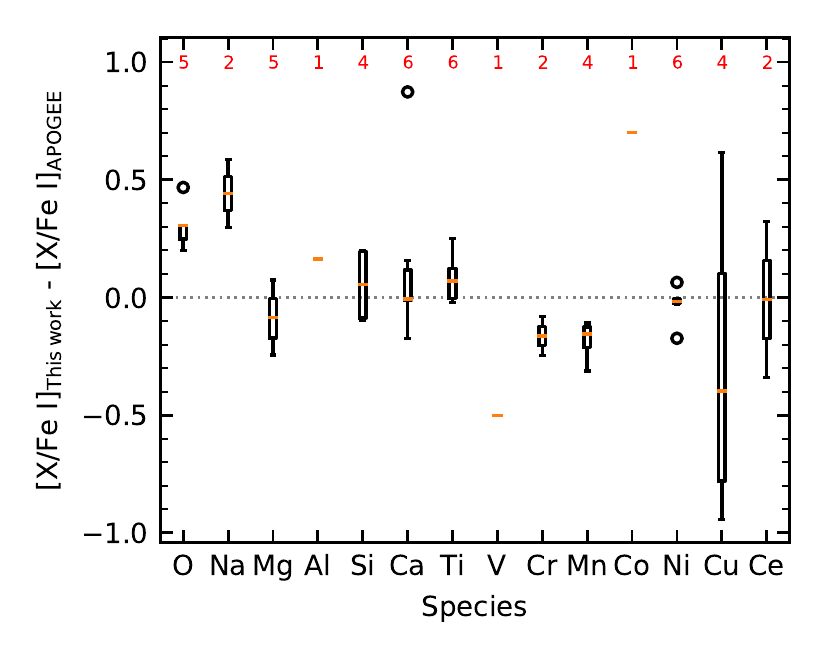}
    \caption{Boxplot showing the abundance ratio differences for each species between this work and APOGEE \citep[][]{Jonsson:2020}. Boxes represent the interquartile ranges (IQR), orange lines are the medians, whiskers correspond to 1.5 $\times$ IQR, and circles are outliers. Red figures on top show the number of data points for each species (i.e., stars that have measurements in both works).
    }
    \label{fig:apogee_comp}
\end{figure}

As briefly discussed in Section~\ref{sec:intro}, information from spectra, astrometry, and stellar oscillations allow an in-depth characterisation of individual stars. In this Section we will both revisit the discussion on the age-chemistry-kinematics relationships of the objects in common with \citet{Epstein:2014} and those picked from \citet{Casagrande:2014aa}.

Six of the seven stars in this study have results from APOGEE \citep[][]{Jonsson:2020}. In Fig.~\ref{fig:apogee_comp} we compare our abundance ratio results for the species with measurements in common with their work. This time we normalise the abundances by \ion{Fe}{I} to make the comparison with APOGEE consistent. Among the [X/Fe] ratios with at least two measurements compared, [O/Fe] and [Na/Fe] have larger values in this study, while the [Cu/Fe] difference among both works has a very large scatter. The outlier in [Ca/Fe] corresponds to KIC\,12017985, whose [Ca/Fe] in APOGEE is $-$0.61~dex. The median differences\footnote{This work values minus APOGEE values.} in T$_{\mathrm{eff}}$, log~g, and [\ion{Fe}{I}/H] are 27~K, $-$0.038~dex, and 0.10~dex, respectively. The median absolute deviations for differences in these parameters are, respectively, 43~K, 0.028~dex, and 0.11~dex. When taking into account the uncertainties from this work and the typical ones from APOGEE ($\sim$100~K in T$_{\mathrm{eff}}$, 0.1~dex in log~g, and 0.02~dex in [\ion{Fe}{I}/H]), the differences in atmospheric parameters are negligible. This also suggests a good accuracy in the seismic-calibrated spectroscopic log~g calculations from \citet{Jonsson:2020}.

\subsection{Revisiting the metal-poor stars from APOKASC}

Five of the metal-poor objects from APOKASC studied by \citet{Epstein:2014} were also analysed in this work, and here we adopt the same identification as in their work for these targets (see Table~\ref{tab:tab1}). From our GBM all five are metal-poor stars on the RGB, and three of them (KIC\,7191496, KIC\,8017159 and KIC\,11563791) have retrograde orbits. KIC\,4345370 was classified by \citeauthor{Epstein:2014} as belonging to the thick disc -- its orbit is the least eccentric in this subset although it shows some moderate departure from the disc plane (|Z$_{\mathrm{max}}$| $\approx$ 2.8 kpc). The modelling of KIC\,12017985 shows a prograde albeit highly eccentric and inclined orbit (|Z$_{\mathrm{max}}$| $\approx$ 4 kpc). However, the results for KIC\,12017985 must be interpreted with caution due to its large RUWE value published in Gaia EDR3.

\begin{figure}
	\includegraphics[width=\columnwidth]{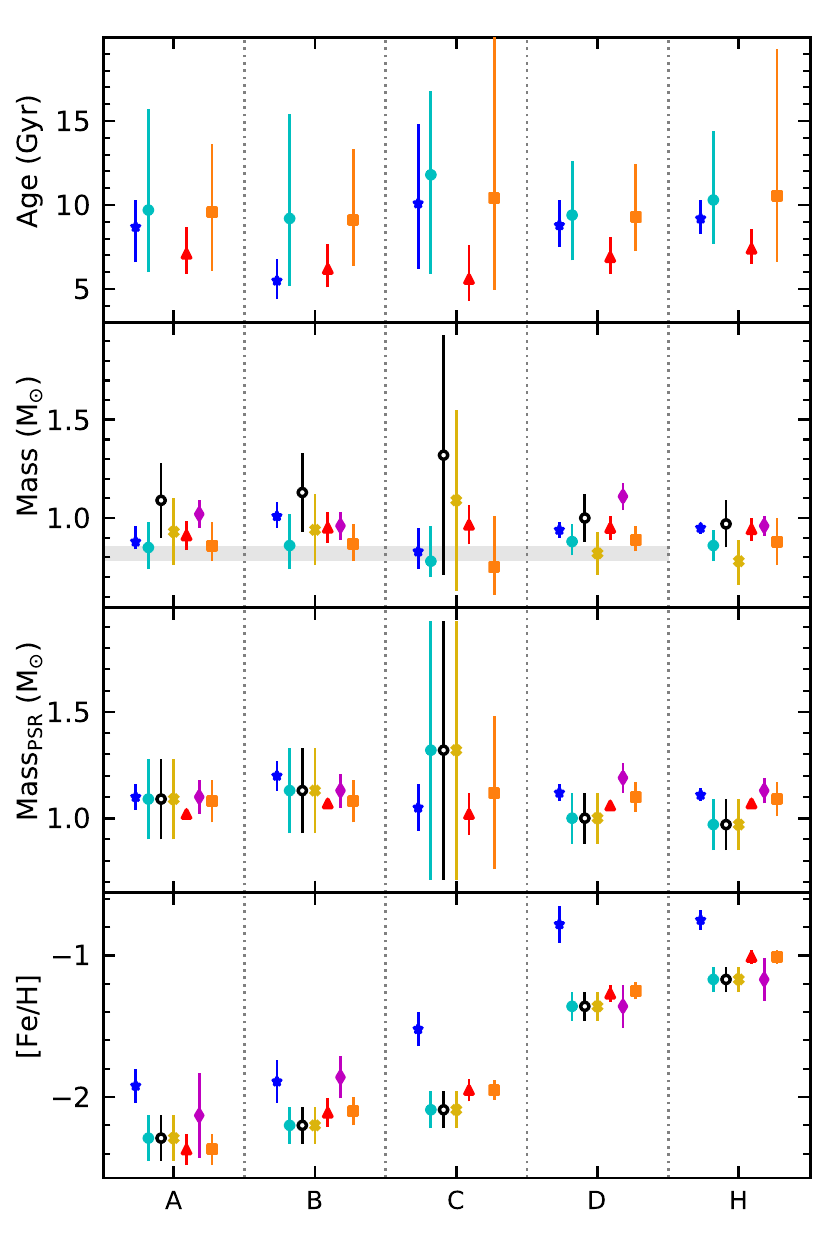}
    \caption{Comparison with values in selected works from the literature for the stars in common with \citet{Epstein:2014}.
    See Table~\ref{tab:tab1} for the corresponding KIC IDs.
    From top to bottom: ages, published masses, masses from \emph{pure scaling relations} given by the published values of $\Delta \nu$, $\nu_{\mathrm{max}}$ and T$_{\mathrm{eff}}$ in each paper, and [Fe/H].
    Blue stars: this work (s$_n$).
    Cyan circles: grid-based modelling using the s$_n$ input set (T$_{\mathrm{eff}}$, [M/H], $\Delta\nu$, $\nu_{\mathrm{max}}$, evolutionary phase, 2MASS $K_s$) with $\Delta \nu$, $\nu_{\mathrm{max}}$ and T$_{\mathrm{eff}}$ published by \citeauthor{Epstein:2014}.
    Black open circles: values published in \citeauthor{Epstein:2014}.
    Golden crosses: \citet[][fig.~12]{Sharma:2016}.
    Red triangles: APOKASC-2 \citep{Pinsonneault:2018}.
    Magenta diamonds: values from \citet{Yu:2018}.
    Orange squares: \citet{Valentini:2019}.
    The grey shaded area represents the 2-$\sigma$ mass range expected by \citeauthor{Epstein:2014}.
    }
    \label{fig:epstein_comp}
\end{figure}

\citet{Epstein:2014} highlighted a few possibilities to explain the discrepancy between their expected mass values for the metal-poor halo stars and their measurements. The authors suggested that a metallicity-based correction would be needed in the scaling relations in order to calculate accurate stellar parameters. Indeed, as shown in Fig.~\ref{fig:epstein_comp}, mass estimations using model-based corrections resulted in lower masses, both in this work and in the literature. Overall, the corrected masses are in or close to the range that \citeauthor{Epstein:2014} have expected for halo stars, the exceptions in this work being KIC\,11563791, whose error bar is $\sim$ 0.03 M$_{\odot}$ above the estimated value in fig.~2 from \citeauthor{Epstein:2014}, and KIC\,12017985.

The ages derived with the s$_n$ input set have their error bars for KIC\,7191496 and KIC\,11563791 (A and D) barely touching the 10 Gyr mark defined as the lower limit for the halo by \citeauthor{Epstein:2014}. Nonetheless, if we consider the older ages derived with the s$_p$ (DR2) input set (see Fig.~\ref{fig:ages_and_plx}(a)), their values for KIC\,7191496 and KIC\,11563791 are more compatible with halo expectations. However, when EDR3 parallaxes are employed the result is qualitatively the same as that obtained with s$_n$ -- i.e., they are younger than the 10-Gyr limit. Meanwhile, the result for KIC\,4345370 (H) is consistent with the age for the thick disc (8-13.77 Gyr). The large error bars for KIC\,8017159 (C) in Fig.~\ref{fig:epstein_comp} might be explained by some difficulty to extract the average seismic parameters is its case, as already discussed by \citeauthor{Epstein:2014}. The fractional $\Delta \nu$ uncertainty adopted here is $\sim$~2~\%, while for all the other four targets in this subset the fractional uncertainties for the same observable are below 1~\%. Also, being on the edge of the parameter space might result in loss of accuracy, as the 0.05 $\mu$Hz difference in $\Delta \nu$ between \citeauthor{Epstein:2014} and the APOKASC-2 value adopted in this work propagates to a 34~\% shift in stellar mass in the scaling relations. A comprehensive discussion on the precision and accuracy of $\Delta \nu$ and $\nu_{\mathrm{max}}$ measurements can be found, e.g., in section 3 of \citet{Yu:2018}. The ages derived in this work for this subset of targets seem to be in an intermediary position between those published by APOKASC-2 and those expected for a halo population -- the exception being KIC\,12017985 (B), whose ages agree in both datasets. For KIC\,8017159, the age uncertainties are larger than those published by APOKASC-2 despite similar error bar sizes in mass.

In order to investigate potential systematics arising from different choices of scaling relation inputs, Fig.~\ref{fig:epstein_comp} also compares masses derived using pure scaling relations (mass$_{\mathrm{PSR}}$), i.e., without any correction factors. For instance, KIC\,12017985 has mass$_{\mathrm{PSR}}$ larger than the uncorrected mass published by \citeauthor{Yu:2018} due only to the larger T$_{\mathrm{eff}}$ adopted in this work, because the average seismic parameters are the same in both studies. Also, it can be seen that the larger uncertainties in both $\Delta \nu$, $\nu_{\mathrm{max}}$ and T$_{\mathrm{eff}}$ in \citeauthor{Epstein:2014} result in much larger error bars in mass$_{\mathrm{PSR}}$. The larger corrected mass of KIC\,7191496 in \citeauthor{Yu:2018} might be a result of a different choice of correction factor, while the larger KIC\,11563791 mass in their results might also be a combination of T$_{\mathrm{eff}}$ and correction factor choices. A similar point about these differences has been made by \citet{Valentini:2019}, whose results are also shown in Fig.~\ref{fig:epstein_comp}.

The cyan points in Fig.~\ref{fig:epstein_comp} show our results if we adopt the values from \citeauthor{Epstein:2014} for our s$_n$ input. As expected from what is shown in the mass$_{\mathrm{PSR}}$ plot in Fig.~\ref{fig:epstein_comp}, the error bars become so large that it is impossible to make any claim about halo membership from these ages. The results from \citet{Valentini:2019} are compared as well. They show larger error bars similar to those from our re-analysis of \citeauthor{Epstein:2014}, possibly linked to their larger input uncertainties. Although both results are compatible with those from the s$_n$ set, the difference in the error bars hint at the required level of precision in the input parameters for age estimation. Also, it is important to note that the reanalysis of \citeauthor{Epstein:2014} has differences between some inputs and fitted parameters for all targets in GBM. Most notably T$_{\mathrm{eff}}$, whose fitted values differ from the inputs by more than 1-$\sigma$ in all five stars, even with input T$_{\mathrm{eff}}$ possessing relatively large uncertainties (125-176~K). The fitted temperatures are always hotter than the inputs. Differences between the results from \citet{Sharma:2016} and our reworked analysis of \citet[][cyan circles in Fig.~\ref{fig:epstein_comp}]{Epstein:2014} are likely the result of different choices of corrections for the scaling relations.

The bottom plot in Fig.~\ref{fig:epstein_comp} shows the different values of metallicity adopted. Although not present in the pure scaling relations, the metallicity plays a role in model-dependent calculations of the stellar parameters -- correction factors and stellar model fitting. Our [M/H] values are systematically higher than those estimated by \citet{Valentini:2019}, although within the error bars. Nevertheless, the effect of different [M/H] choices seems to be mild. For instance, adopting \citeauthor{Valentini:2019} or \citeauthor{Epstein:2014} [M/H] values would increase the mass of KIC\,7191496 by 0.02 or 0.04 M$_{\odot}$, respectively, without affecting the resulting uncertainties. The resulting age would be qualitatively the same, with differences from our adopted value staying below 1-$\sigma$.

\subsubsection{KIC\,12017985}

The age of 5.5$^{+1.3}_{-1.1}$~Gyr, estimated for KIC\,12017985, seems too young for Halo membership, being closer to what is expected for the thin disc. However, assuming that its Gaia EDR3 astrometric solution can be trusted for modelling its kinematics, the dynamical characteristics of KIC\,12017985 are not expected for a thin disc star (e.g., relatively low L${z}$ and large eccentricity). Our reanalysis of the \citet{Epstein:2014} data and the results from \citet{Valentini:2019} give older ages, but their associated uncertainties are too large and reach values compatible with both thin disc and the age of the Universe. From the chemical point of view, KIC\,12017985 abundances are compatible with those expected from an old halo population -- e.g., [Mg/Fe] = [Ca/Fe] = 0.36~dex, and it has high \ion{Ti}{I} and O abundances as well. Some of its s-process elements abundances are relatively high (e.g., La, Ce and Nd). KIC\,12017985 has detection of Li; A(Li)$_{\mathrm{LTE}}$ = 1.23~dex.

\citet{Bandyopadhyay:2020} made a high-resolution chemical analysis of KIC\,12017985 and their abundance ratios are systematically lower than ours with very few exceptions (Ni and Ba). The differences are probably due to different sets of atmospheric parameters adopted - their classical spectroscopic analysis gave values closer to our initial spectroscopic guesses for the atmospheric parameters, however our adopted IRFM T$_{\mathrm{eff}}$ is 275~K hotter than the \citeauthor{Bandyopadhyay:2020} value and our seismic log~g is $\sim$ 0.7~dex larger. If our adopted T$_{\mathrm{eff}}$ and metallicity values are replaced by the purely spectroscopic ones, \texttt{BASTA} is not able to find a solution, as it pushes the results towards the boundary of the parameter space (age $\approx$ 19 Gyr). Also, no agreement is found between input and output T$_{\mathrm{eff}}$, $\Delta \nu$ and $\nu_{\mathrm{max}}$ when the spectroscopic parameters are employed.

\citet{Belokurov:2020} argue, using Gaia DR2 data, that large RUWE values are related to binarity. If KIC\,12017985 (RUWE = 2.8 in Gaia EDR3) is a member of a multiple system, its relatively large mass could be explained by past mass accretion from a companion that it is masking the age if this red giant, i.e., KIC\,12017985 would be a blue straggler. The results found to date for this object make it a good target for a long-term campaign of radial velocity monitoring in order to check if it is a member of a binary system. Another possibility is KIC\,12017985 being a member of some uncatalogued dwarf galaxy recently accreted by the Milky Way.

\subsubsection{KIC\,4345370 and KIC\,11563791}

KIC\,11563791 and KIC\,4345370 display similar metallicities ([Fe/H] $\sim$ -0.75). However, while the later is a member of the thick disc, the former has a retrograde orbit. KIC\,11563791 is more metal-rich than the other stars with retrograde orbits in this study, and it presents a relatively peculiar composition. It presents near-solar Mg and Ca, while at the same time it is Si- and O-rich (see Table~\ref{tab:abundances1}). It is also an outlier in Y, with [Y/Fe] = 0.44, and it has low [Eu/Fe] relative to objects of similar metallicity in this work.

Among the stars with Zn measurements, KIC\,11563791 and KIC\,4345370 display strong enhancement in this species, around +0.7~dex as shown in Table~\ref{tab:abundances1}. They also display (relative) depletion in Cr and Mn, which could hint at an Hypernova-dominated origin of their primoridal gas \citep{Nomoto:2013aa}. However, the Hypernova scenario would be challenged by their observed V and Co abundances, as both elements are expected to be enriched together with Zn on sites with very large explosion energies. Another caveat is the fact that such a scenario is predicted in a more metal-poor regime, as also discussed by \citeauthor{Nomoto:2013aa}. Given their ages ($\sim$9~Gyr), Zn production in the Universe is thought to be dominated by core-collapse supernovae at the chronological point of their formation. However, this pair of stars also display mild enhancement in Ba and La abundances (see Table~\ref{tab:abundances1}), and their [Ba/Fe] abundances suggest enrichment mostly by the s-process \citep[][]{Kobayashi:2020}.

\subsection{KIC\,4671239 and KIC\,6279038}

KIC\,4671239 and KIC\,6279038 are metal-poor objects identified by the SAGA Survey \citep{Casagrande:2014aa}. The $\Delta \nu$ values published by \citeauthor{Casagrande:2014aa} and APOKASC-2 for KIC\,6279038 are larger than the $\Delta \nu$ published by \citeauthor{Yu:2018} by 0.136~$\mu$Hz, or 3-$\sigma$. The difference in $\nu_{\mathrm{max}}$ is not as large, being lower than 1-$\sigma$. Such difference is expected to have a significant impact in age estimation, although not in log~g.

\begin{figure}
	\includegraphics[width=\columnwidth]{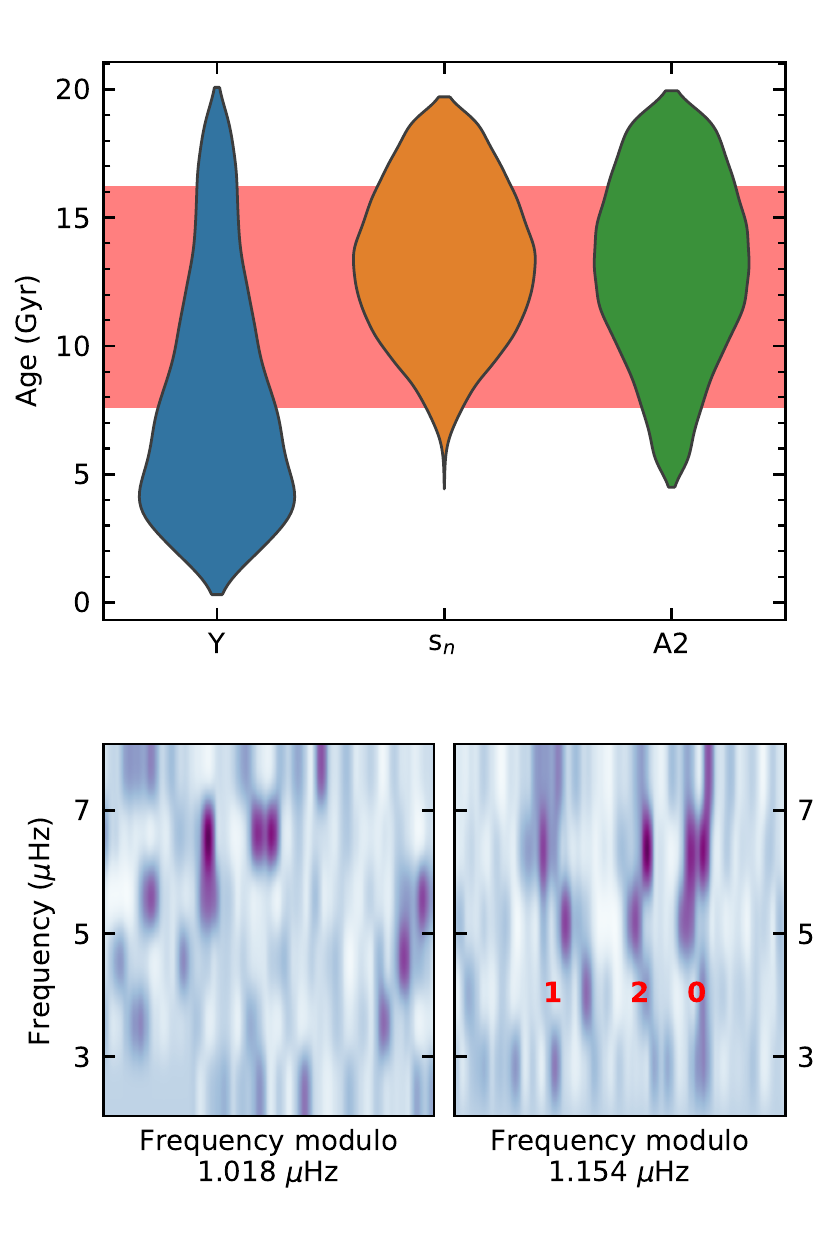}
    \caption{\emph{Top}: Violin plots illustrating the age distributions for KIC\,6279038 resulting from different values of input parameters -- T$_{\mathrm{eff}}$, [M/H], $\Delta \nu$, and $\nu_{\mathrm{max}}$.
    \emph{Left}: using $\Delta \nu$ and $\nu_{\mathrm{max}}$ from \citet{Yu:2018} (Y), and T$_{\mathrm{eff}}$ and [M/H] calculated in this work;
    \emph{centre}: from the values shown in Tables \ref{tab:tab1} and \ref{tab:atmparam};
    \emph{right}: using input parameters published by \citet[][APOKASC-2]{Pinsonneault:2018} (A2).
    The shaded area shows the 1-$\sigma$ age interval given in the APOKASC-2 catalogue.
    \emph{Bottom}: echelle plots folded by $\Delta \nu$ values published in APOKASC-2 (right) and \citeauthor{Yu:2018} (left). Numbers in the right plot indicate the angular degrees \emph{l}.
    }
    \label{fig:violin_627}
\end{figure}

In order to evaluate the impact of the 3-$\sigma$ $\Delta \nu$ deviation in age estimations, new s$_n$ runs were made employing the observables published by \citeauthor{Yu:2018} and APOKASC-2. The resulting distributions are shown in the violin plots in Fig.~\ref{fig:violin_627}. It is evident from inspection that the increased value of $\Delta \nu$ -- adopted in this work -- yields a much larger age. It must be pointed out that the disparity between different sets of input values taken from the literature is large enough to characterise this metal-poor object as either young or old -- despite the considerable uncertainties in the 'young' case. Visual inspection of the echelle plots created with the corresponding $\Delta \nu$ values as frequency modulo indicate that the value published in APOKASC-2 is more accurate than that from \citeauthor{Yu:2018}. The ridges expected to be seen in the echelle plots of solar-like oscillators are more apparent in the former. Also, the mode structure seen in the (correct) echelle plot in Fig.~\ref{fig:violin_627} agrees with the expected (and observed) behaviour for luminous stars, with the dipole modes being further away from the l=0,2 pair at higher frequencies \citep[][figs. 1 and 3]{Stello:2014}.

The disagreement between the results for KIC\,6279038 illustrates the shortcomings of the characterization of low log~g stars using Kepler data. Its Gaia parallax gives no improvement to the estimation, because it is the most distant star in this work and parallaxes from both data releases are similar -- $\varpi_{\mathrm{DR2}}$~=~0.1928~$\pm$~0.0271~mas after employing the zero-point correction from \citet{Zinn2019}, while (corrected) $\varpi_{\mathrm{EDR3}}$~=~0.1898~$\pm$~0.0099 mas \citep{GaiaEDR3:2021,Lindegren:2021}. 

KIC\,4671239 is a lower RGB (log g $\sim$ 3) object and the most metal-poor target studied in this work. Few species were measured due to low S/N in the observed spectrum. However, it is an interesting object as its [Mg/Fe] and [Ca/Fe] are slightly above 0.1 dex -- although it shows enhancement in Ti -- while being a relatively massive and metal-poor object, with 1.01~$\pm$~0.02~M$_{\odot}$ and [Fe/H] = -2.63~$\pm$~0.20. It has a very elongated polar orbit according to our modelling (J$_{\phi}$/J$_{\mathrm{tot}}$ $\approx$ 0.3; R$_{\mathrm{peri}}$ = 1.0~kpc; eccentricity = 0.79), and, kinematically, it seems to belong to the heated thick disc \citep{Koppelman:2018,Helmi:2020}.

The mass/age of KIC\,4671239 is too large/young when compared with expectations for a metal-poor halo star. The simplest explanation is the possibility of this object being a blue straggler, as suspected for KIC\,12017985. However, at least from its Gaia EDR3 astrometry, there is no evidence of binarity. On the other hand, KIC\,4671239 seems to have a peculiar photometry: a spectral energy distribution fitting with VOSA \citep{Bayo:2008aa} using the atmospheric parameters from Table~\ref{tab:atmparam} reveals some flux excess on the WISE W3 region \citep[$\sim$~12~$\mu$m,][]{Cutri:2014}. Unfortunately, its W4 magnitude only has an upper limit published (quality flag 'U'), but it can be speculated that KIC\,4671239 did suffer a merger in the past, and this infrared excess would be the observational evidence of a debris disc resulting from such event. The same argument has been made by \citet{Yong:2016aa} to suggest that a few young $\alpha$-rich stars displaying infrared excess in their photometry would indeed be blue stragglers, and it can explain why the asteroseismic mass of this red giant is too large when its low metallicity is taken into account in the context of Galactic chemical evolution. The presence of two out of seven blue stragglers in our sample of metal-poor red giants would not be a surprise given that around one fifth of nearby halo stars are thought to be blue stragglers \citep{Casagrande:2020}.

\subsection{Accreted stars?}

\begin{figure*}
	\includegraphics{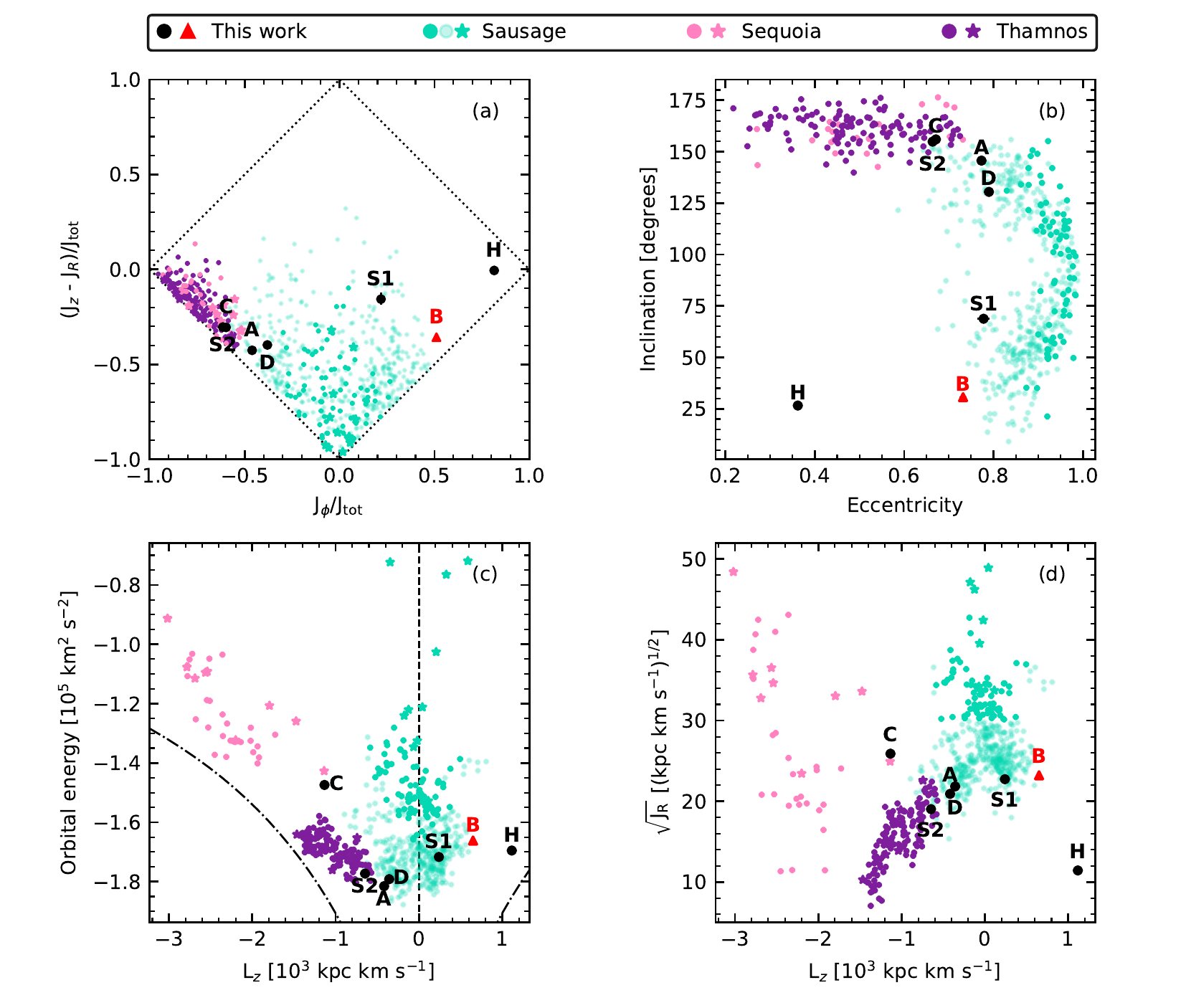}
    \caption{Dynamical parameter space evaluated in this work.
    Black circles represent the stars in this study other than KIC\,12017985 (B) (Gaia EDR3 RUWE = 2.8), marked as a red triangle. See Table~\ref{tab:tab1} for the corresponding KIC IDs. Error bars not shown are smaller than their respective markers.
    Star markers are data from \citet{Monty:2020}.
    Points are from \citet{Yuan:2020}.
    Turquoise colour corresponds to Gaia-Sausage, pink refers to Gaia-Sequoia, purple represents low-energy retrograde stars corresponding to Thamnos (i.e., Sequoia~G1 from \citeauthor{Monty:2020}; DTGs 21, 24 and 29 from \citeauthor{Yuan:2020}).
    Light turquoise markers represent stars identified as Gaia-Sausage by \citeauthor{Yuan:2020} that do not pass more strict constraints seen in the literature (see text).
    (a) Action space.
    (b) Orbital inclination as function of eccentricity.
    (c) Orbital energy vs. z-component of angular momentum L$_z$. The black dashed line (L$_z$ = 0) corresponds to a non-rotating orbit. The dash-dotted lines show circular (prograde and retrograde) orbits.
    (d) Square root of the radial action J$_R$ as function of L$_z$.
    }
    \label{fig:dynamics}
\end{figure*}

Given the intriguing results in stellar ages derived from asteroseismology for the sample analysed in this work -- mostly still younger than usually expected for halo stars, it might be insightful to check their chemodynamics. As noted by \citet{Koppelman:2019}, the highly retrograde overdensity originally called Gaia-Sequoia \citep{Barba:2019,Myeong:2019} can be separated in two distinct groups in the energy-angular momentum space -- high- and low-energy groups. The former would be \emph{Sequoia proper} while the later was called Thamnos by \citeauthor{Koppelman:2019}. In Fig.~\ref{fig:dynamics}(c) their separation is located at E/E$_{\odot}$~$\sim$~1. 

\begin{figure}
	\includegraphics{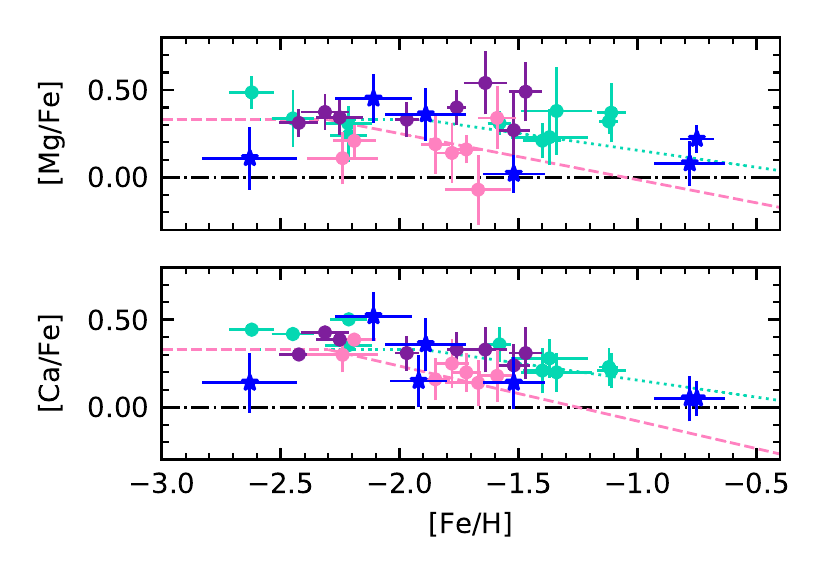}
    \caption{[Mg,Ca/Fe] plots showing Gaia-Sequoia and Gaia-Sausage $\alpha$-knee fits according to \citet[][pink and turquoise lines, respectively]{Monty:2020}.
    \emph{Blue stars}: this work.
    From \citeauthor{Monty:2020} -- \emph{Pink and purple circles}: Gaia-Sequoia high (E/E$_{\odot}$ < 1) and low orbital energy samples, respectively.
    \emph{Turquoise circles}: Gaia-Sausage.
    }
    \label{fig:haloplot}
\end{figure}

\citet{Monty:2020}, in their study of accreted populations, suggested that the $\alpha$-knee (the location at which the [$\alpha$/Fe] ratio begins to drop) of Gaia-Sequoia \emph{high orbital energy} starts at [Fe/H] $\approx$ -2.2. Their results are reproduced in Fig.~\ref{fig:haloplot}, as well as in Fig.~\ref{fig:dynamics}(a,c,d). When $\alpha$-element composition is taken into account one can spot certain agreement between KIC\,8017159 ([Fe/H] = -1.52) and the Gaia-Sequoia $\alpha$-knee fits in [Mg,Ca/Fe]. The low-energy group (i.e., Thamnos, purple circles in Fig.~\ref{fig:haloplot}) seems to have its $\alpha$-knee located at a higher metallicity with respect to the high-energy group. Looking at Fig.~\ref{fig:dynamics}(c), KIC\,8017159 (C) falls in the valley that seems to divide both groups (Sequoia and Thamnos), sharing the same region of the Energy-Momentum space as well as similar orbital parameters with the globular cluster FSR~1758 \citep{Myeong:2019}. If we consider the group of stars identified with the Gaia-Sausage event by \citet{Yuan:2020}, which overlaps with Gaia-Sequoia in Fig.~\ref{fig:dynamics}(a,b) it could be argued that KIC\,8017159 belongs to Gaia-Sausage instead of Gaia-Sequoia/Thamnos. On the other hand, if we apply constraints similar to those adopted by \citet{Feuillet:2020} and \citet{Limberg:2021} to the Gaia-Sausage event (eccentricity $\geq$ 0.8; -600 $\leq$ J$_{\phi}$ $\leq$ +500 kpc km s$^{-1}$; 900 $\leq$ J$_R$ $\leq$ 2500 kpc km s$^{-1}$), KIC\,8017159 clearly does not belong to it. Its non-membership of Gaia-Sausage seems clear when we take into consideration the radial action in Fig.~\ref{fig:dynamics}(d).

Meanwhile, it can be suggested that KIC\,6279038 (S2) belongs to Thamnos. Its metallicity places it between both Sequoia and Sausage $\alpha$-knees, and combining its $\alpha$-enhancement and position in the Energy-Momentum space makes it a typical member of Thamnos. If Sequoia and Thamnos belong to the same accreted satellite, KIC\,6279038 would be an older, more metal-poor member, while KIC\,8017159 would represent a younger, less metal-poor member that formed just before the accretion. However, if the more broad dynamical definition of \citeauthor{Yuan:2020} is considered, it could either be a member of Thamnos or Gaia-Sausage (see Fig.~\ref{fig:dynamics}), as both groups are not separable by chemical composition in Fig.~\ref{fig:haloplot}.

From Fig.~\ref{fig:dynamics}, KIC\,7191496 and KIC\,11563791 (A and D) seem to lie near the transition zone between Gaia-Sausage and Thamnos. They might be members of the former if the \citeauthor{Yuan:2020} definition is considered. However, KIC\,7191496 has [Ca/Fe] more compatible with Gaia-Sequoia, putting it in the low-$\alpha$ sequence in Fig.~\ref{fig:haloplot}.

Regarding KIC\,11563791, it is more metal-rich than the other objects in this study with retrograde orbits, and its chemical composition is similar to KIC\,4345370, a thick disc member. The lower [Mg/Fe] of the former with respect to the later may hint on an \emph{ex-situ} origin \citep{DiMatteo:2019}. Also, the [Ni/Fe] in KIC\,11563791 is 0.2 dex lower than in KIC\,4345370, further suggesting an accreted origin, as objects formed in satellites tend to be deficient in this element \citep{Montalban:2021}. However, the Ni abundance in KIC\,7191496 is solar, and its [Ca/Fe] seems too low in comparison with the trend from Gaia-Sausage in Fig.~\ref{fig:haloplot}, as mentioned in the previous paragraph. In this case, KIC\,7191496 and KIC\,11563791 would be unrelated in Galactic history despite showing very similar ages and dynamics. Such result is not unexpected, as over-densities of multiple accretion events may overlap between each other and with the \emph{in-situ} population \citep{Jean-Baptiste:2017}. Given that the degeneracy among accretion events expected in dynamical space is broken only partially with the use of two out of three observational sources of information (chemistry, kinematics, chronology), precise red giant ages derived from solar-like oscillations for a \emph{large number} of halo stars, as done in this work for a small dataset, may be useful to further characterise different events (if separated in time), thus increasing our understanding of the violent early stages of the Galaxy.

\section{Conclusions}
\label{sec:conclusions}

In this study we took advantage of the availability of high-quality astrometric, spectroscopic and asteroseismic data in order to explore how confidently we can apply asteroseismic inferences of metal poor stars for Galactic Archaeology. Our findings can be summarised as follows:

\begin{itemize}
    \item Red giant ages from grid-based modelling are very sensitive to high-precision parallaxes such as those published by the Gaia Collaboration, at least in the parallax range represented in our sample. Hence, inaccuracies in parallax zero-point may result in shifts of several Gyr in modelled stellar ages, as the best fitting solutions may correspond to inaccurate distance moduli, resulting in an incorrect position in the HR~diagram. For red giants with precise determinations of $\Delta \nu$ and $\nu_{\mathrm{max}}$, the trade-off between accuracy and precision in age may result in a loss of accuracy when high-precision parallaxes are included, unless a very accurate calibration has been obtained for the parallax zero-point.
    
    \item However, the grid-based modelling results are not particularly sensitive to the code/grid adopted, at least for those tested here -- \texttt{BASTA}/BaSTI and \texttt{PARAM} with models from \citet{Rodrigues:2017}. The choice of mass-loss parameter (zero mass-loss or $\eta =$ 0.3) also did not have any significant effect in the results. However, this is not expected for red clump stars where mass-loss is relevant \citep{Casagrande:2016aa}.
    
    \item \citet{Epstein:2014} were correct when pointing out that the pure asteroseismic scaling relations overestimate masses on the RGB, as already confirmed by several studies like those mentioned in Section~\ref{sec:intro}. However, as seen in Fig.~\ref{fig:epstein_comp}, the uncertainties resulting from grid-based modelling using the observables adopted by \citet{Epstein:2014} and the corrections on the scaling relations still do not allow us to draw qualitative conclusions. In this work we confirm the required level of precision in the average asteroseismic parameters needed for APOKASC-2-level age precision using grid-based modelling. However, the differences in the results from APOKASC-2 and this work seen in Fig.~\ref{fig:epstein_comp} still need to be better understood. It is worth noting that the adoption of different metallicity values has a very mild effect in resulting masses and ages.
    
    \item There is a good amount of evidence for arguing that KIC\,12017985 may be an evolved blue straggler. This object is a good target for long-term radial velocity monitoring, to check the possibility of a past mass transfer event that masked its initial mass, as well as to confirm (or rule out) that its large RUWE value published by Gaia is linked to binarity. From our results, only a mass transfer event from a former AGB companion can be ruled out, as there is no detection of extreme s-process enhancement. KIC\,4671239 is an evolved blue straggler candidate as well, due to its low metallicity and young age, and there is tentative evidence of infrared excess, which might be the telltale sign of a relatively recent merger in this system. At the same time, their membership of unknown dwarfs recently accreted cannot be ruled out.
    
    \item KIC\,6279038 exemplifies the sensitivity of the modelled fundamental parameters to the asteroseismic observables. Different published values of $\Delta \nu$ resulted in ages that would radically change the interpretation on the nature of this object. These results help to highlight the importance of a very careful inspection of the oscillation power spectra in a detailed analysis. This object would be useful to characterise the accreted population of Thamnos if it is confirmed as a member, being one of its oldest stars.
    
    \item Apart from KIC\,4345370 and (maybe) the two blue straggler candidates, all stars are likely members of the accreted halo. Of those, only KIC\,6279038 does not fit the age distribution of Gaia-Sausage \citep{Montalban:2021}, as well as being a potential member of Thamnos. A study with an expanded dataset that includes asteroseismic data for Sequoia and Thamnos could indicate the age distributions of these overdensities and clarify the status of KIC\,8017159.
    
\end{itemize}

\section*{Acknowledgements}

We thank the anonymous referee for important comments and suggestions. LC is the recipient of an ARC Future Fellowship (project number FT160100402). Parts of this research were conducted by the Australian Research Council Centre of Excellence for All Sky Astrophysics in 3 Dimensions (ASTRO~3D), through project number CE170100013. We also want to thank Zhen Yuan and GyuChul Myeong for kindly providing supporting data shown in Fig.~\ref{fig:dynamics}. This research made use of \texttt{astropy}, \url{http://www.astropy.org}, a community-developed core \texttt{pyhton} package for Astronomy \citep{astropy:2013,astropy:2018}, \texttt{numpy} \citep{numpy:2020}, \texttt{matplotlib} \citep{Matplotlib:2007}, \texttt{seaborn} \citep{seaborn:2021}, \texttt{scipy} \citep{scipy:2020}, \texttt{Lightkurve}, a \texttt{python} package for Kepler and TESS data analysis \citep{lightkurve:2018}, and \texttt{echelle} \citep{echelle:2020}. Some of the data presented herein were obtained at the W. M. Keck Observatory, which is operated as a scientific partnership among the California Institute of Technology, the University of California and the National Aeronautics and Space Administration. The Observatory was made possible by the generous financial support of the W. M. Keck Foundation. The authors wish to recognize and acknowledge the very significant cultural role and reverence that the summit of Maunakea has always had within the indigenous Hawaiian community.  We are most fortunate to have the opportunity to conduct observations from this mountain. This publication makes use of data products from the Two Micron All Sky Survey, which is a joint project of the University of Massachusetts and the Infrared Processing and Analysis Center/California Institute of Technology, funded by the National Aeronautics and Space Administration and the National Science Foundation.

\section*{Data Availability}

The data underlying this article are available in the article and online through supplementary material.



\bibliographystyle{mnras}
\typeout{}
\bibliography{references} 




\appendix

\section{Some extra material}

\begin{table}
	\centering
	\caption{Internal uncertainties of the atmospheric parameters from Table~\ref{tab:atmparam}, calculated as described in Section~\ref{sec:unc}. The last column also shows the S/N for the spectrum of each target.}
	\label{tab:apdx_atmunc}
	\begin{tabular}{lrrrrr}
		\hline
		Star & $\sigma$T$_{\mathrm{eff}}$ & $\sigma$log g & $\sigma$[M/H] & $\sigma v_t$ & S/N \\
		             & K & dex & dex & km s$^{-1}$ & @590nm \\
        \hline
         KIC4345370 &  71 & 0.004 & 0.04 & 0.04 & 130 \\
         KIC4671239 & 145 & 0.002 & 0.11 & 0.09 &  57 \\
         KIC6279038 &  79 & 0.020 & 0.06 & 0.11 &  85 \\
         KIC7191496 &  96 & 0.007 & 0.08 & 0.18 &  75 \\
         KIC8017159 &  58 & 0.003 & 0.05 & 0.14 & 165 \\
        KIC11563791 &  56 & 0.005 & 0.03 & 0.09 &  95 \\
        KIC12017985 &  90 & 0.007 & 0.06 & 0.06 & 170 \\
        \hline
	\end{tabular}
\end{table}

\begin{table*}
	\centering
	\caption{Internal uncertainties, in dex, of the 'absolute' abundance A(X) for each species X.}
	\label{tab:apdx_axunc}
	\begin{tabular}{lrrrrrrrrrrrrr}
		\hline
               Star &    O &   Na &   Mg &   Al &   Si &   Ca &   Sc &   Ti &    V &   Cr &   Mn &   FeI & FeII  \\
	    \hline
         KIC4345370 & 0.06     & 0.11     & 0.08     & 0.08     & 0.02     & 0.11 & 0.06     & 0.09 & 0.13     & 0.06     & 0.10     & 0.11 & 0.07     \\ 
         KIC4671239 & $\cdots$ & $\cdots$ & 0.07     & $\cdots$ & $\cdots$ & 0.10 & $\cdots$ & 0.19 & $\cdots$ & $\cdots$ & $\cdots$ & 0.20 & $\cdots$ \\ 
         KIC6279038 & 0.03     & $\cdots$ & 0.06     & $\cdots$ & $\cdots$ & 0.12 & 0.02     & 0.17 & 0.13     & $\cdots$ & 0.14     & 0.17 & 0.16     \\ 
         KIC7191496 & $\cdots$ & $\cdots$ & $\cdots$ & $\cdots$ & $\cdots$ & 0.17 & 0.04     & 0.19 & $\cdots$ & $\cdots$ & $\cdots$ & 0.21 & 0.12     \\ 
         KIC8017159 & 0.03     & $\cdots$ & 0.05     & $\cdots$ & 0.03     & 0.16 & 0.07     & 0.26 & $\cdots$ & $\cdots$ & 0.10     & 0.18 & 0.13     \\ 
        KIC11563791 & 0.04     & 0.10     & 0.11     & $\cdots$ & 0.08     & 0.09 & 0.06     & 0.15 & 0.10     & 0.04     & 0.10     & 0.12 & 0.15     \\ 
        KIC12017985 & 0.11     & 0.05     & 0.06     & $\cdots$ & 0.03     & 0.13 & 0.08     & 0.15 & 0.16     & $\cdots$ & 0.09     & 0.17 & 0.17     \\ 
        \hline
		            &   Co &   Ni &   Cu &   Zn &   Sr &    Y &   Zr &   Ba &   La &   Ce &   Nd &   Sm &   Eu \\
		\hline
		 KIC4345370 & 0.06     & 0.14     & 0.08     & 0.07     & 0.12     & 0.17     & 0.16     & 0.14     & 0.08     & 0.17     & 0.20     & 0.05     & 0.03     \\
		 KIC4671239 & $\cdots$ & $\cdots$ & $\cdots$ & $\cdots$ & $\cdots$ & $\cdots$ & $\cdots$ & $\cdots$ & $\cdots$ & $\cdots$ & $\cdots$ & $\cdots$ & $\cdots$ \\
		 KIC6279038 & $\cdots$ & 0.09     & 0.15     & 0.04     & $\cdots$ & 0.08     & $\cdots$ & 0.05     & 0.10     & 0.03     & 0.07     & 0.13     & $\cdots$ \\
		 KIC7191496 & $\cdots$ & 0.12     & $\cdots$ & 0.05     & $\cdots$ & 0.09     & $\cdots$ & 0.05     & $\cdots$ & $\cdots$ & $\cdots$ & $\cdots$ & $\cdots$ \\	   
		 KIC8017159 & $\cdots$ & 0.08     & $\cdots$ & 0.11     & $\cdots$ & 0.07     & $\cdots$ & $\cdots$ & 0.01     & 0.04     & 0.06     & $\cdots$ & 0.07     \\	   
		KIC11563791 & $\cdots$ & 0.13     & 0.07     & 0.22     & $\cdots$ & 0.06     & $\cdots$ & 0.15     & 0.04     & $\cdots$ & 0.10     & 0.03     & 0.05     \\	   
		KIC12017985 & $\cdots$ & 0.08     & 0.09     & 0.04     & $\cdots$ & 0.08     & $\cdots$ & 0.06     & 0.09     & 0.05     & 0.12     & $\cdots$ & 0.02     \\	   
	    \hline
	\end{tabular}
\end{table*}


\bsp	
\label{lastpage}
\end{document}